\documentclass{aa}

\usepackage[varg]{txfonts}
\usepackage{natbib}
\bibpunct{(}{)}{;}{a}{}{,} % to follow the A&A style

\begin{document}

\title{A new trigger mechanism for coronal mass ejections}
\subtitle{The role of confined flares and photospheric motions \\ in the formation of hot flux ropes}

\author{A. W. James\inst{1,}\inst{2} %[0000-0001-7927-9291]
\and L. M. Green\inst{2} %[0000-0002-0053-4876]
\and L. van Driel-Gesztelyi\inst{2,}\inst{3,}\inst{4} %[0000-0002-2943-5978]
\and G. Valori\inst{2}} %[0000-0001-7809-0067]

\institute{European Space Agency (ESA), European Space Astronomy Centre (ESAC), Camino Bajo del Castillo, s/n, 28692 Villanueva de la Ca{\~n}ada, Madrid, Spain
\email{alexander.james@esa.int}\label{inst1}
\and Mullard Space Science Laboratory, University College London, Holmbury St. Mary, Dorking, Surrey, RH5 6NT, UK
\and LESIA, Observatoire de Paris, Universit\'e PSL, CNRS, Sorbonne Universit\'e, Universit\'e Paris Diderot, 5 place Jules Janssen, 92190 Meudon, France
\and Konkoly Observatory, Research Centre for Astronomy and Earth Sciences, Budapest, Hungary}

\date{Received 29 June 2020 / Accepted 20 October 2020}

\abstract
%Context
{Many previous studies have shown that the magnetic precursor of a coronal mass ejection (CME) takes the form of a magnetic flux rope, and a subset of them have become known as `hot flux ropes' due to their emission signatures in $\sim$10 MK plasma.}
%Aims
{We seek to identify the processes by which these hot flux ropes form, with a view of developing our understanding of CMEs and thereby improving space weather forecasts.}
%Methods
{Extreme-ultraviolet observations were used to identify five pre-eruptive hot flux ropes in the solar corona and study how they evolved. Confined flares were observed in the hours and days before each flux rope erupted, and these were used as indicators of episodic bursts of magnetic reconnection by which each flux rope formed. The evolution of the photospheric magnetic field was observed during each formation period to identify the process(es) that enabled magnetic reconnection to occur in the $\beta<1$ corona and form the flux ropes.}
%Results
{The confined flares were found to be homologous events and suggest flux rope formation times that range from 18 hours to 5 days. Throughout these periods, fragments of photospheric magnetic flux were observed to orbit around each other in sunspots where the flux ropes had a footpoint. Active regions with right-handed (left-handed) twisted magnetic flux exhibited clockwise (anticlockwise) orbiting motions, and right-handed (left-handed) flux ropes formed.}
%Conclusions
{We infer that the orbital motions of photospheric magnetic flux fragments about each other bring magnetic flux tubes together in the corona, enabling component reconnection that forms a magnetic flux rope above a flaring arcade. This represents a novel trigger mechanism for solar eruptions and should be considered when predicting solar magnetic activity.}


\titlerunning{A new trigger mechanism for CMEs} %less than 60 characters
\authorrunning{A.W. James et al.}

\maketitle

%-----------

\section{Introduction} \label{sec:intro}

% Definition of CMEs
Coronal mass ejections (CMEs) are expulsions of magnetised plasma from the solar atmosphere and represent the most energetic events in the Solar System. These eruptions are ultimately the result of a magnetic energy storage and release process that is the consequence of the emergence of magnetic flux from the solar interior combined with photospheric motions that further evolve the coronal field. The free magnetic energy stored in the corona is in the form of field-aligned electric currents, and it is this energy that powers CMEs. 

% CME triggers and drivers: introductory comments
Separate processes are required to describe the slow build-up of electric currents in the corona before a CME and the rapid onset and acceleration of the eruption. Following the convention of \citet{aulanier2010formation}, in this work, CME initiation mechanisms are classified as either triggers or drivers. 
Triggers create a stable structure before an eruption and the driving process is responsible for the fast expansion phase of the eruption. 
A wide range of mechanisms have been identified as CME triggers, including sunspot rotation (\citealp[e.g.][]{yan2012sunspot,torok2013initiation,james2017on-disc}), magnetic reconnection (e.g. flux cancellation; \citealp{van1989formation, yardley2018cancellation}, and tether-cutting; \citealp{moore2001onset}), and the helical kink instability of a flux rope \citep{hood1979kink,torok2005confined}. 
There are two main groups of theories pertaining to how the rapid expansion of a CME is driven. 
One group assumes that CMEs are driven by reconnection, which may or may not involve a flare, for example, in the `breakout' scenario \citep{antiochos1999model,temmer2010combined,karpen2012mechanisms}.
The other group assumes that CMEs are driven by an ideal magnetohydrodynamic (MHD) instability or loss of equilibrium involving a flux rope, such as the torus instability \citep{vanTend1978development,kliem2006torus,demoulin2010criteria}.
For a summary of triggers and drivers, see Table 1 of \citet{green2018eruptions}.

A substantial number of observational studies have been carried out to test the theories that involve flux ropes present at the time of eruption. Early observational studies tested the flux cancellation mechanism of \cite{van1989formation} whereby a low altitude flux rope forms via magnetic reconnection in the photosphere or chromosphere. This low altitude flux rope may have a so-called bald patch separatrix (BPS) configuration in which the underside of the rope (that runs along the photospheric polarity inversion line; PIL) is rooted in the dense, high plasma $\beta$ environment of the lower atmosphere, as shown by \cite{green2009flux}. Such ropes can be formed in both emerging active regions in which opposite polarity fragments collide, or in decaying active regions where the separation, natural fragmentation, and dispersal of emerging bipoles can bring polarities into contact with neighbouring polarities of opposite sign in sufficiently complex active regions. 

More recently, extreme-ultraviolet (EUV), X-ray, and radio observations have shown evidence of pre-eruption flux ropes containing 10 MK plasma (known as hot flux ropes) and that have their underside in the low plasma $\beta$ environment of the corona. These flux ropes are seen best when they are close to the solar limb \citep[e.g.][]{reeves2011atmospheric,patsourakos2013direct,Nindos2015common}. 
Hot flux ropes also appear to be formed as a consequence of magnetic reconnection, but this time in the corona, where the reconnection not only leads to the formation of the flux rope, but also a confined flare \citep{patsourakos2013direct,james2017on-disc}.

The formation of a flux rope in the corona necessitates quasi-separatrix layers (QSLs) separating the rope from the external field it is embedded within.
\cite{james2018model} showed that a hot flux rope had a specific magnetic configuration known as a hyperbolic flux tube \cite[HFT;][]{titov2002hft}. In an HFT, two intersecting QSLs are present and an X-line runs along the underside of the rope. 
In this configuration, perturbations to the coronal field caused by, for example, photospheric flows may preferentially lead to reconnection in the current sheet beneath the flux rope \citep{demoulin1996qslcurrent}. This reconnection could build additional magnetic flux in to the flux rope over time, and perhaps even play a role in the eruption of the rope.

In order to fully understand the physics of CMEs and be able to predict them, relevant triggers must be identified so that pre-eruptive structures can be recognised, and knowledge of whether an eruption of these structures can be successfully driven must also be obtained. However, most observational studies that identified higher altitude coronal flux ropes at the limb, which are likely to be of the HFT configuration, were unable to observe the evolution of the photosphere beneath the forming flux ropes, hindering the identification of the trigger mechanisms involved in these CMEs.

This work follows on from two studies that investigated an HFT rope that formed and erupted near disc centre allowing both the photospheric and coronal evolution to be tracked (\citeauthor{james2017on-disc} \citeyear{james2017on-disc}; hereafter Paper I, and \citeauthor{james2018model} \citeyear{james2018model}; hereafter Paper II). 
The combination of observations and modelling in Papers I and II showed the presence of an HFT flux rope with its axis $\sim 0.2\ \textrm{R}_{\sun}$ above the photosphere. 
In this event, the flux rope formed as a consequence of magnetic reconnection in the corona, driven as photospheric orbiting motions of flux fragments about each other led to a collision of flux systems. 
While magnetic reconnection in the photosphere associated with the formation of BPS ropes injects photospheric plasma into the structure \citep{baker2013plasma}, spectroscopic measurements in Paper I showed that the studied flux rope had a coronal plasma composition, supporting the conclusion that the orbiting motions had indeed driven magnetic reconnection in the corona transforming a sheared arcade into an HFT rope. 

As well as likely playing a key role in flux rope formation, orbiting motions of magnetic flux systems about each other can inject magnetic energy and helicity in to the arcade field that overlies the rope causing the field to inflate and weakening its confining effect on the underlying flux rope \citep{torok2013initiation}.
This scenario is consistent with the findings of Paper II that suggested the flux rope eruption was driven by the torus instability due to the magnetic field gradient above the rope reaching a critical value.
The motion of magnetic flux fragments around each other may also be framed in the broader observational context of sunspot rotation (spin). The causal connection between rotating sunspots and the onset of solar activity has been well-studied and linked to the formation of sigmoids, flaring, and CMEs (\citealp[e.g.][]{gerrard2003rotation,brown2003rotating,yan2012sunspot,torok2013initiation,vemareddy2016rotation}).
The sunspots in each of these studies feature multiple umbrae within one rotating sunspot, and may therefore be similar to the `orbiting' scenario of eruption triggering described above.

Papers I and II represent a case-study of a single event. A natural extension of these previous works is to investigate whether this process is observed in other eruptive active regions and could therefore be viewed as a CME trigger mechanism.
In this study, we investigate the formation of five HFT flux ropes in four active regions, including the active region studied in Paper I and Paper II. Photospheric orbiting motions are quantified in each active region to identify whether the same processes that formed the flux rope in Papers I and II are systematic in HFT flux rope formation.
The criteria used to select events are given in Sect. \ref{sec:sci3_data_and_events}.
Sections \ref{sec:method_orbit}, \ref{sec:method_height}, and \ref{sec:method_chirality} describe the methods used to quantify photospheric orbiting motions, estimate the height of coronal structures, and determine the chiralities of each active region, respectively.
The observations of each eruption and the measured motions are detailed in Sect. \ref{sec:sci3_observatons}, the results are interpreted and discussed in Sect. \ref{sec:sci3_discussion}, and conclusions are presented in Sect. \ref{sec:sci3_conclusions}.

\section{Data and event selection} \label{sec:sci3_data_and_events}

%Event selection
Events with HFT flux ropes present in the corona before erupting as CMEs were selected from the era of the \textit{Solar Dynamics Observatory} (SDO; \citealp{pesnell2012SDO}).
Observational signatures of HFT flux ropes include, for example, sigmoids and plasmoids with underlying arcade fields \citep{reeves2011atmospheric}. 
Arcades can be identified in many of the EUV passbands of the SDO's \textit{Atmospheric Imaging Assembly} (AIA; \citealp{lemen2012atmospheric}), but sigmoids and plasmoids appear only in those bands with the hottest temperature responses $\sim$ 10 MK.
The 131 \AA{} channel of AIA was used to identify sigmoids and plasmoids in the corona because it has a peak in temperature response at $\sim10^{7.0}\ \textrm{K}$, however, it also has another peak at the lower temperature of $10^{5.6}\ \textrm{K}$ \citep[see Fig. 11 of][]{boerner2012initial}.
This means AIA 131 \AA{} images contain contributions from both relatively hot and cold plasma.
To ensure that only hot plasma signatures were identified, images from the relatively cool 171 \AA{} channel (temperature response peak at $10^{5.8}\ \textrm{K}$) were used for comparison. 
Any feature that appeared in the 131 \AA{} images but did not appear in the 171 \AA{} channel was assumed to correspond to hotter, $\sim10^{7.0}\ \textrm{K}$ plasma, and therefore may be a flux rope signature. 

Additionally, observations from the 193 \AA{}, and 211 \AA{} channels of AIA (that image coronal plasma at $\sim10^{6.2}\ \textrm{K}$ and $\sim10^{6.3}\ \textrm{K}$ respectively) were used to identify EUV dimmings, and the 1600 \AA{} channel (that images plasma in the lower solar atmosphere at $\sim10^{5.0}\ \textrm{K}$) was used to locate flare ribbons. 
EUV dimmings and flare ribbons manifest at the footpoints of erupting flux ropes \citep{janvier2014electric}, and are therefore critical for determining the locations where the photospheric motions associated with flux rope formation within the orbiting scenario should be studied.
AIA data were processed to level 1.5 using the aia\_prep routine available in SolarSoft. 

The CMEs, and the solar hemisphere from which they originated, were identified using images from the \textit{Large Angle and Spectrometric Coronagraph} (LASCO; \citealp{brueckner1995large}) onboard the \textit{Solar and Heliospheric Observatory} (SOHO; \citealp{domingo1995soho}) and the \textit{Solar Terrestrial Relations Observatory} (STEREO; \citealp{kaiser2008stereo}) coronagraphs.
The STEREO spacecraft were $\sim \pm 120\degr$ from the Sun-Earth line at the times of the selected events (at first in 2012, and then having approximately swapped positions with each other by 2017, albeit with only STEREO A still functioning), giving complementary perspectives of the CMEs.

%HMI
The evolution of the photospheric flux and horizontal motions in each CME source region were studied using data from the \textit{Helioseismic and Magnetic Imager} (HMI; \citealp{scherrer2012hmi}) onboard SDO.
White-light continuum images and radial magnetic field observations were obtained from the SHARP data series (Spaceweather HMI Active Region Patch; \citealp{bobra2014sharp}). 
In events where the SHARP magnetograms contained more than one region of strong magnetic flux, such as multiple active regions, the data were cropped to contain only the desired source active region. Pixels with magnetic flux densities lower than $30\ \textrm{G}$ were excluded from the flux calculations to reduce the effect of noise.
In order to study the evolution of the photosphere before the eruptions, observations of the photospheric magnetic field over a number of days were analysed. The radial magnetic field measurement is increasingly affected by noise outside $\pm 60\degr$ from disc-centre \citep{liu2012comparison}, so only CME source regions that evolved within this longitude range were used.
The heliographic latitudes and longitudes of the studied active regions were referenced from the Debrecen Photoheliographic Data sunspot catalogue\footnote{\url{http://fenyi.solarobs.csfk.mta.hu/en/databases/DPD/}}.
Furthermore, because flux cancellation is not expected during the formation of HFT flux ropes, the radial field magnetograms were used to confirm that the active regions selected in this study did not exhibit significant flux cancellation along the CME-source polarity inversion lines.

%GOES
Full-Sun integrated X-ray light curves were obtained from the \textit{Geostationary Operational Environmental Satellite} \textit{X-ray Sensor} (GOES XRS) system and used in combination with EUV images from SDO/AIA to identify flares and brightenings in the CME source active regions. 

In summary, the selection criteria for events with hot flux ropes were: a white light CME was clearly identifiable in coronagraph data; the CME source region was on the Earth-facing side of the Sun within $\pm 60\degr$ of disc-centre; a sigmoid or a plasmoid was observed before eruption in an imaging waveband that senses plasma at a temperature of $\sim 10^{7}\ \textrm{K}$; an arcade of loops was present under the identified flux rope before eruption; and no significant magnetic flux cancellation was observed along the polarity inversion line associated with the CME.

Using these criteria led to the selection of five CMEs from four active regions.
The dates of the five CMEs are given in  Table \ref{tbl:P3_events}, along with information about the source active regions.
In each of these four active regions, magnetic flux fragments were observed to orbit each other during the days before the CMEs.

\begin{table}
    \caption{CME source regions.}
 	\centering
		\begin{tabular}{ccccc}
            \hline\hline
            Date of CME & NOAA AR & Lat. & Lon. & Hale \\
			\hline
			13 Mar 2012 & 11429 & N18$\degr$ & W59$\degr$ & No  \\
            13 Jun 2012 & 11504 & S17$\degr$ & E23$\degr$ & Yes \\
            14 Jun 2012 & 11504 & S17$\degr$ & E10$\degr$ & Yes \\
            08 Oct 2012 & 11585 & S19$\degr$ & W33$\degr$ & Yes \\
            14 Jul 2017 & 12665 & S6$\degr$  & W34$\degr$ & Yes \\
			\hline
		\end{tabular}
 		\tablefoot{
 		Heliographic latitudes (Lat.) and longitudes (Lon.) of the studied active regions are given at the time of the eruption onset. 
 		The fifth column indicates whether or not each active region conforms to Hale's polarity law.
 		}
		\label{tbl:P3_events}
\end{table}

\section{Methods} \label{sec:sci3_methods}

\subsection{Quantifying the orbital motion of emerging magnetic flux fragments} \label{sec:method_orbit}

This study builds on that presented in Paper I, in which the orbital motion of a magnetic flux fragment around a sunspot was estimated by-eye using two white-light continuum images that were taken 24 hours apart. However, in this work, a more systematic method is used to track the motion of individual magnetic flux fragments, as explained in this section.

\begin{figure}
 \centering
 \resizebox{\hsize}{!}{\includegraphics{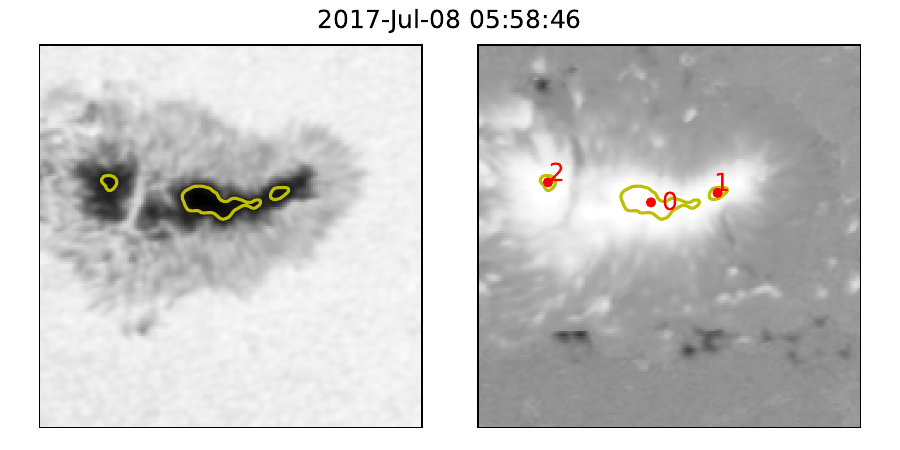}}
 \caption{Identifying magnetic flux fragments. Left: Contours are drawn at fixed intensity levels on HMI continuum images to enclose individual umbral fragments. Right: The continuum contours are overlaid on the radial component of magnetic field. Positive polarity field is shown in white, and negative in black. The flux-weighted centroid of each fragment is shown in red and is assigned a unique identifying number.}
 \label{fig:contour_masks}
 \end{figure}

HMI continuum images from the SHARP data series were used to study the motion of magnetic flux fragments in the photosphere.
The full passage of each active region across the solar disc was followed at an image cadence of 6 hours.
Contours were set on each HMI SHARP continuum image to encircle individual umbral fragments, and the flux-weighted centre of each fragment was found using the radial magnetic field strength inside each contour (see Fig. \ref{fig:contour_masks}).
Across different active regions and at different times during the evolution of a given active region, different contour values were chosen between 10000--25000 DN/s in the continuum images to best isolate specific fragments.

\begin{figure}
\centering
  \resizebox{\hsize}{!}{\includegraphics{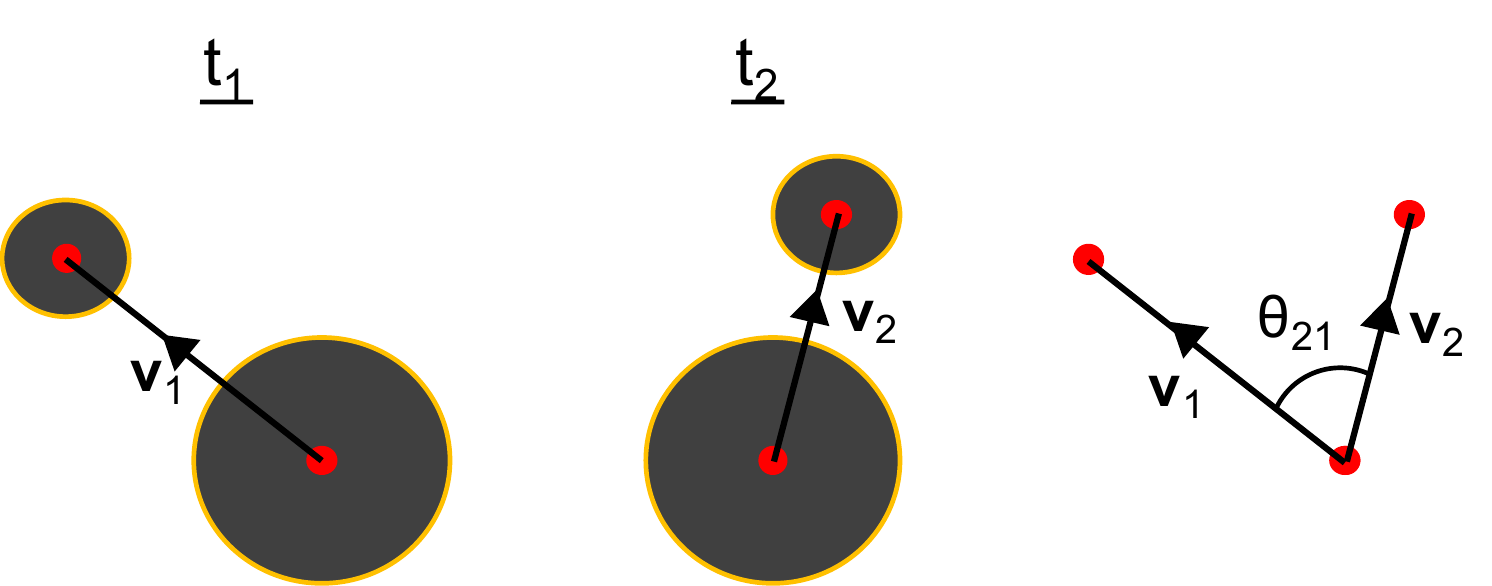}}
    \caption{Illustration of the method used to track the relative orbital motion of magnetic flux fragments.}
    \label{fig:orbit_method}
\end{figure}

In the first image of a sequence (at time $t_{1}$), two fragments were selected: one as the `central' fragment and one as the `orbiting' fragment, and a vector ($\textbf{v}_{1}$) was drawn between their flux-weighted centroids.
In the next image in the sequence (at time $t_{2}$), the orbiting fragment had moved relative to the central fragment, and a new vector ($\textbf{v}_{2}$) was drawn between the newly-positioned flux-weighted centroids.
By comparing the vectors in successive images (illustrated in Fig. \ref{fig:orbit_method}), the angle the orbiting fragment had moved around the central fragment by was calculated using the relation
\begin{equation}
    \theta_{21} = \arccos{ \frac{\textbf{v}_{1} \cdot \textbf{v}_{2}}{\left|\textbf{v}_{1}\right|\left|\textbf{v}_{2}\right|} \,.}
\label{eqn:theta_dot_product}
\end{equation}
This gives $0 \le \theta_{21} \le 180$, and therefore does not distinguish between rotation in the clockwise and anticlockwise directions.
In order to obtain a signed orbit angle, $-180 \le \theta_{21} \le 180$, the relations
\begin{equation}
\begin{array}{l}
    \left(\textbf{v}_{1} \times \textbf{v}_{2}\right) \cdot \textbf{v}_{n} \;  < \; 0 \;\;\; \rightarrow \;\;\; 0  < \theta_{21} <  180 \, \textrm{, and} \\ \left(\textbf{v}_{1} \times \textbf{v}_{2}\right) \cdot \textbf{v}_{n} \;  > \; 0 \;\;\; \rightarrow \; -180  < \theta_{21} <  0
\label{eqn:theta_sign_cross}
\end{array}
\end{equation}
were used, where $\textbf{v}_{n}$ is the vector normal to $\textbf{v}_{1}$ and $\textbf{v}_{2}$, oriented positively out of the image plane towards the observer.
Here, the convention is chosen that clockwise (anticlockwise) motions correspond to positive (negative) orbit angles.
This process was repeated for every image in the time sequence to track the orbital motion over time.

In most events, separate pairs of fragments were tracked at different times to best quantify the observed motions in each evolving active region.
The vectors between different pairs of tracked fragments are colour-coded in, for example, Fig. \ref{fig:rot_2012march}, and the corresponding orbit rates of different fragments are colour-coded in, for example, Fig. \ref{fig:rot_goes_flux_201203}.

At some time steps, it was not possible for the tracking routine to distinguish two fragments, for example after they had merged together.
However, the seemingly-merged sunspots still often seemed to be comprised of distinct fragments, as evidenced by observed light bridges or the later re-separation in to similar fragments.
In these cases, the solid-body rotation of the merged sunspot was quantified as a proxy for how the previously-individual fragments continued to move around each other.
This was done by linearly fitting a vector to the major axis of the elliptical sunspot umbra to best describe the orientation of the sunspot and following the rotation of this vector through time in the same way as described by Equations \ref{eqn:theta_dot_product} and \ref{eqn:theta_sign_cross}.
The effectiveness of this method is limited in the case of merged sunspots that are approximately circular.

It is difficult to quantify error in the orbit rates. The primary source of uncertainty arises from the computed flux-weighted centroids that can move over time as multiple fragments coalesce. This can lead to instances where the orbit rate appears to stop or even reverse direction briefly. It is hard to quantify this effect, but we attempt to account for it by showing three-point moving averages of the measured orbit rates, for example, in Figure \ref{fig:rot_goes_flux_201203}, in order to better show the qualitative trend of the orbiting with less of an effect from frame-to-frame error.

\subsection{Estimating flux rope altitude} \label{sec:method_height}

Determining the heights of the identified flux ropes above the photosphere is important for supporting the interpretation of an HFT configuration and for examining the overall stability of a flux rope. In this work, flux rope heights are estimated from observations using geometrical considerations and assumed symmetry.

The assumption is made that the mid-point of either an observed sigmoid (which is interpreted as representing plasma emission from the underside of a flux rope) or the bottom of a plasmoid lies radially above the centre of a photospheric PIL. That is, that the flux rope is not inclined relative to the perpendicular direction. In reality, this will not necessarily be true, as the symmetry of the system is affected by many factors (e.g. the spatial flux distribution at the photosphere, the distribution of currents in the active region, the dynamical evolution of the sigmoid, etc.). 
The asymmetries and projection effects make it difficult to implement a method to automatically determine the mid-points of the sigmoids, so we chose this point by-eye based on the geometry of the sigmoid. Specifically, we found the middle of the relatively straight central section between the curved ends of the fully-illuminated S-shaped loops.
Due to the nature of the assumptions being made, any resulting heights should be taken as order-of-magnitude estimations only.

From plane-of-sky solar observations, two coordinates were taken: the coordinate of the centre of the coronal sigmoid or plasmoid-base ($x_{\mathrm{cor}},y_{\mathrm{cor}}$) and the point at the middle of the photospheric PIL between sunspots ($x_{\mathrm{pil}},y_{\mathrm{pil}}$).
Then, the height of the coronal point above the photospheric PIL was estimated using either the difference in $x$ coordinates or the difference in $y$ coordinates as
\begin{equation}
    h_{x} = \left(\frac{x_{\mathrm{cor}}}{x_{\mathrm{pil}}} - 1\right)R_{\sun}, \\
    h_{y} = \left(\frac{y_{\mathrm{cor}}}{y_{\mathrm{pil}}} - 1\right)R_{\sun} \,.
\label{eqn:hx_hy}
\end{equation}
If the assumption that the coronal point lies radially above the photospheric point is correct, then $h_{x} = h_{y}$, but otherwise these two equations will give different height estimates.
Whenever this method was applied in this work, both $h_{x}$ and $h_{y}$ were computed and the average was taken to give a qualitative estimate of height.
Errors in $h_{x}$ and $h_y$ were calculated from the uncertainty in selecting the coordinates of the PIL and sigmoid centres.

\subsection{Chirality determination} \label{sec:method_chirality}

As part of investigating the formation of the hot flux ropes, the chirality of each active region was determined in two ways.
Firstly, the chirality of magnetic flux emerging in to the photosphere was inferred from observations of magnetic tongues \citep{lopez-fuentes2000counterkink,luoni2011twisted} in radial field HMI observations.
If the leading polarity in a bipolar emerging active region extends to the south (north) of the trailing polarity, the emerging magnetic flux has right-handed (left-handed) twist.

Secondly, the chirality of the coronal magnetic field was indicated by observations of forward-S or reverse-S sigmoids in EUV observations from the 131 \AA{} channel of AIA.
Sigmoids with forward-S (reverse-S) shapes are a signature of magnetic field with right-handed (left-handed) twist in the corona \citep{longcope1998twist}.

\section{Observations and results} \label{sec:sci3_observatons}

\subsection{13 March 2012}

\subsubsection{Coronal evolution} \label{sec:mar2012coronal}
 
 \begin{figure} 
 \centering{}
 \resizebox{\hsize}{!}{\includegraphics{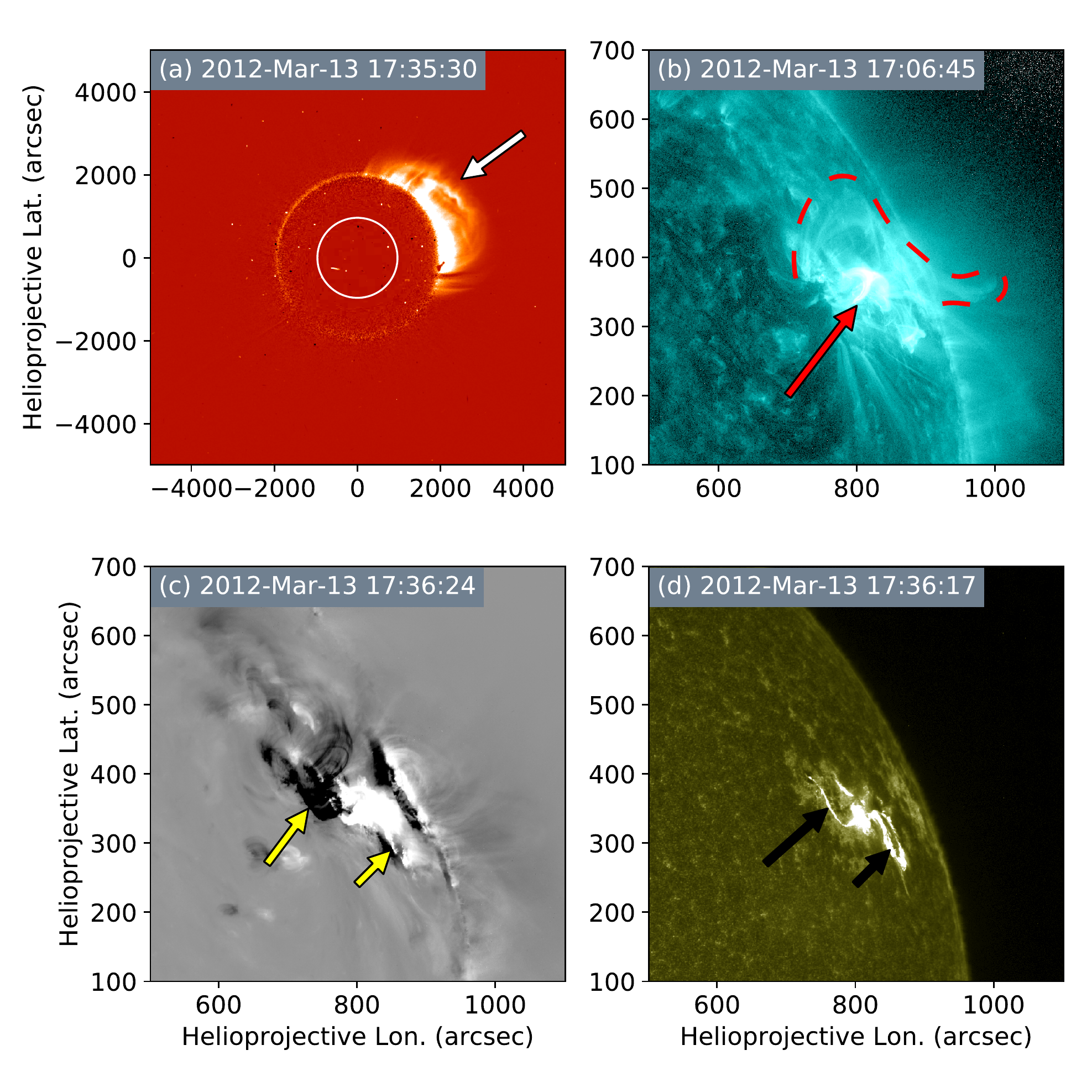}}
 \caption{Observations of the 13 March 2012 eruption. (a) A white-light CME (indicated by the white arrow) seen by LASCO C2. (b) AIA 131 \AA{} image of a sigmoid (highlighted by a dashed red line) brightening above a flaring arcade (indicated by a red arrow). (c) AIA 211 \AA{} base difference image showing twin EUV dimmings (marked by two yellow arrows). (d) AIA 1600 \AA{} image showing hooked flare ribbons (indicated by two black arrows).}
 \label{fig:euv_2012march} 
 \end{figure}

On 13 March 2012, a CME and accompanying M7.9 flare occurred in NOAA AR 11429 at $\sim$17:10 UT. The CME was first observed in white-light by LASCO C2 at 17:36 UT (Fig. \ref{fig:euv_2012march}a) when the active region was $59\degr$ west of central meridian. An inverse-S (left-handed) sigmoid and underlying arcade brightened in the active region at 16:30 UT, shortly before the onset of the eruption (see Fig. \ref{fig:euv_2012march}b).
The ends of the sigmoid were rooted in the east and west parts of the active region, and twin EUV dimmings and hooked flare ribbons were observed in these locations during the eruption itself (Figures \ref{fig:euv_2012march}c and \ref{fig:euv_2012march}d respectively). 
The centre of the pre-eruptive sigmoid (at 16:47 UT) is estimated to be at an altitude of 56 $\pm$ 5 Mm above the photosphere, however, this is found by averaging two rather different estimates, with $h_{x}$ = 91 $\pm$ 8 Mm and $h_{y}$ = 22 $\pm$ 3 Mm.
The relatively large difference between $h_{x}$ and $h_{y}$ in this estimate may be due to the strong projection effects arising from the active region's position nearly 60$\degr$ west of disc-centre.
Due to the lack of overlap of the estimated heights, the height estimate here should be treated with caution. This is the only event in this study where there is no overlap between the $h_{x}$ and $h_{y}$ estimates.
Together, the observations detailed above suggest the presence of an HFT flux rope with footpoints in the east and west sides of the active region that erupted during the CME.

The previous CME from NOAA AR 11429 occurred three days prior, on 10 March at 17:44 UT, in association with an M8.4 flare.
Similarly to the CME on 13 March, the flare arcade produced during this event on 10 March spanned the full width of the active region, indicating that both eruptions originated from the active region's internal polarity inversion line.
Furthermore, a white-light CME with a bright front and cavity was seen at 18:18 UT in LASCO C2 images, suggesting that the CME had a flux rope structure (\citealp{vourlidas2013cmes} and the references within). 
Assuming that the 10 March CME fully ejected any flux rope that was present in the active region, the flux rope that erupted on 13 March must have formed at some time during the 3 days between eruptions. The specific time(s) of flux rope formation can be inferred by identifying solar flares (and therefore magnetic reconnection episodes) during this time.

GOES detected four flares in NOAA AR 11429 between the CMEs on 10 and 13 March 2012.
Of these four, two flares illuminated arcades that spanned across the whole active region in the same way as the flux rope eruptions and exhibited flare ribbons in the 1600 \AA{} channel of AIA that were in the same locations as were seen during the 13 March CME. Furthermore, hot plasma structures in the AIA 131 \AA{} channel appeared above the flare arcades with shapes similar to -- and in the same location as -- the sigmoid associated with the 13 March CME.
These two flares were a C4.1 flare on 12 March at 22:20 UT and a C3.1 flare on 13 March at 06:55 UT. Observations from the SOHO/LASCO and STEREO coronagraphs confirm these were confined events, supporting the scenario that these flares were associated with the construction, but not eruption, of a flux rope.
These observations suggest the flux rope had begun to form at least 19 hours before the onset of the CME on 13 March 2012.

\subsubsection{Photospheric evolution}

 \begin{figure}
 \centering{}
 \resizebox{\hsize}{!}{\includegraphics{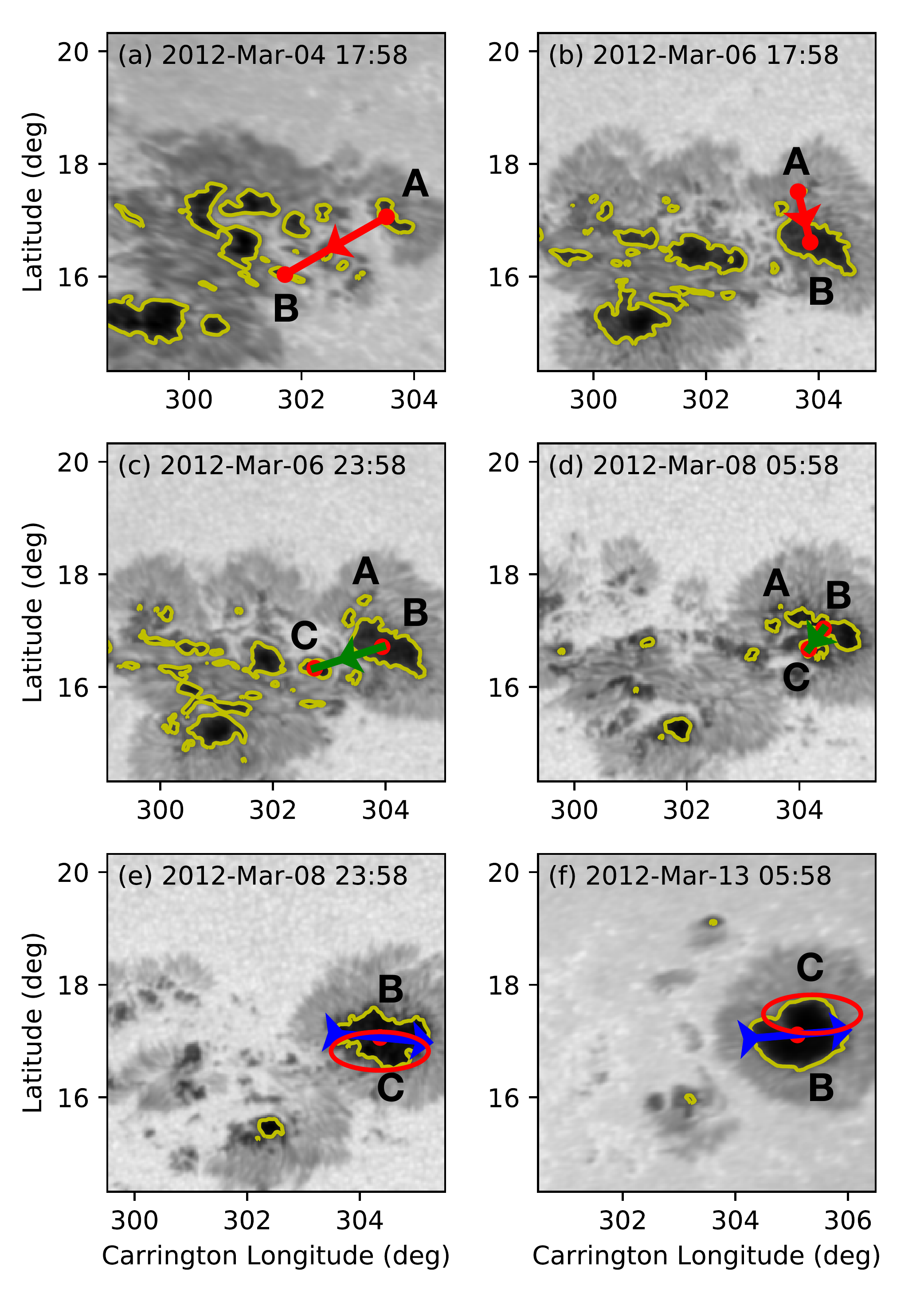}}
 \caption{Anticlockwise motion of newly-emerged flux around the pre-existing positive (leading) sunspot in NOAA AR 11429. Tracked fragments of flux are labelled A--C. In each image, vectors are drawn either to connect the flux-weighted centroids of two orbiting fragments (panels a-d) or to best-fit the major axis of merged fragments (panels e-f). In the bottom two panels, red circles highlight the position of a fragment (C) whose $\approx$180$\degr$ orbit from the south of the sunspot to the north between panels was not successfully tracked using the vector method. This figure is available as an online movie.}
 \label{fig:rot_2012march}
 \end{figure}
 
 \begin{figure} 
 \centering{}
 \resizebox{\hsize}{!}{\includegraphics{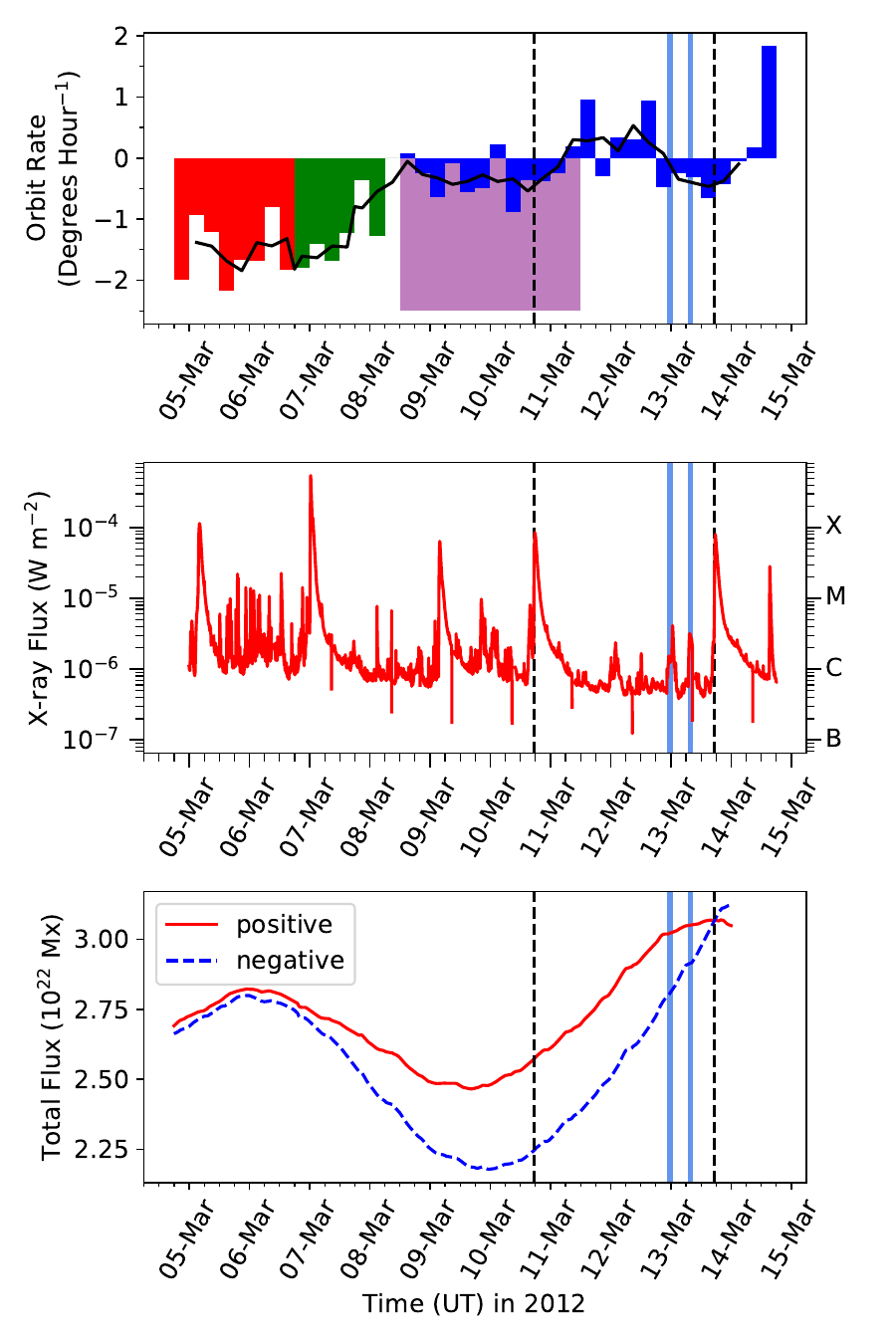}}
 \caption{Top: Measured orbital motion of chosen magnetic flux fragments in 6-hour intervals. The colours correspond to different choices of fragments (see Fig. \ref{fig:rot_2012march}), except for the purple block which corresponds to the average orbiting estimated by-eye that was not detected by the automated method. The black curve represents the orbit angles smoothed using a three-point moving average. Middle: Full-disc integrated GOES soft X-ray lightcurve. Bottom: The evolution of magnetic flux in NOAA AR 11429, made using the radial magnetic field component, $B_{\mathrm{r}}$, of the HMI SHARP data series and smoothed with a 24-hour moving average.
 Vertical dashed lines indicate the times of the CME onsets on 10 and 13 March 2012, and the thin, light-blue vertical bars show the timings of confined flares associated with flux rope formation described in Sect. \ref{sec:mar2012coronal}.}
 \label{fig:rot_goes_flux_201203}
 \end{figure}

NOAA AR 11429 rotated on to the solar disc on 3 March 2012 and contained two sunspots at that time. The leading sunspot had positive polarity and the trailing spot was negative, meaning the active region did not follow Hale's law for a northern-hemisphere region in solar cycle 24.
A new flux emergence episode began on 4 March 2012 at two sites between the pre-existing sunspots. The orientation of the magnetic tongues of this newly emerging flux indicated that it had left-handed twist (negative chirality). 

The positive and negative fragments of the emerging flux moved towards the pre-existing sunspots of the same polarities.
As the fragments approached the sunspots, they orbited around them in an anticlockwise sense.
The strongest orbiting was observed around the leading sunspot, and the EUV dimmings (shown in Fig. \ref{fig:euv_2012march}c) suggest that the flux rope had a footpoint in this leading spot.
Therefore, the leading sunspot was chosen as the location to quantify orbiting motion.

The motion of positive magnetic flux around the leading sunspot was tracked by following distinct fragments using the method described in Sect. \ref{sec:method_orbit}.
Here, we refer to certain fragments of magnetic flux with letters A--C, as in Fig. \ref{fig:rot_2012march}.
At first, emerging fragment B moved towards and anticlockwise around pre-exisiting fragment A, and then fragment C emerged and orbited anticlockwise around fragment B.
The orbiting motion is quantified in Table \ref{tbl:rotation} and also represented in Figures \ref{fig:rot_2012march} and \ref{fig:rot_goes_flux_201203}.
In addition, Fig. \ref{fig:rot_goes_flux_201203} also contains the GOES X-ray lightcurve over the duration of the orbiting and the evolution of the positive and negative magnetic flux of the active region for comparison.

\begin{table*}
		\caption{Orbital motions of umbral flux fragments relative to previously-emerged umbrae and the observed active region chiralities.}
		\centering
		\begin{tabular}{lccccccc} 
            \hline\hline
            NOAA & Start Time & End Time & Total Orbit & Avg. Orbit & Emerging & Coronal & Estimated FR \\
            AR & (UT) & (UT) & ($\degr$) & ($\degr$ day$^{-1}$) & Flux & Field & Height (Mm) \\
			\hline
			11429 & 04 Mar 12 18:00\tablefootmark{a} & 06 Mar 12 18:00 & -74 & -37 & LH & LH & 56 $\pm$ 5 \\
                  & 07 Mar 12 00:00\tablefootmark{a} & 08 Mar 12 06:00 & -36 & -29 \\
                  & 08 Mar 12 12:00\tablefootmark{a} & 14 Mar 12 12:00 & -19 & -3 \\
                  & 08 Mar 12 12:00\tablefootmark{b} & 14 Mar 12 12:00 & -180 & -30 \\
            \hline
			11504 & 09 Jun 12 18:00 & 10 Jun 12 18:00 & +2 & +2 &  
			        RH & RH & 173 $\pm$ 26 \\
                  & 11 Jun 12 06:00 & 12 Jun 12 12:00 & +60 & +48 & & & 89 $\pm$ 30 \\
                  & 12 Jun 12 18:00 & 15 Jun 12 18:00 & +186 & +62 \\
			\hline
			11585 & 03 Oct 12 06:00 & 06 Oct 12 00:00 & -111 & -40 &                 LH & LH & 98 $\pm$ 25 \\
                  & 06 Oct 12 12:00 & 08 Oct 12 00:00 & -41 & -27 \\
                  & 08 Oct 12 12:00 & 09 Oct 12 00:00 & -7 & -14 \\
            \hline
			12665 & 08 Jul 17 00:00 & 10 Jul 17 00:00 & -92 & -46 &                  LH & LH & 71 $\pm$ 20 \\
                  & 10 Jul 17 06:00 & 14 Jul 17 00:00 & -47 & -12.5 \\
                  & 14 Jul 17 00:00 & 15 Jul 17 00:00 & -43 & -43 \\
			\hline
		\end{tabular}
 		\tablefoot{
 		Start Time and End Time delineate periods that orbiting motions of magnetic flux fragments were tracked over. In the total and averaged (avg.) daily orbital motions, positive angles are clockwise and negative angles are anticlockwise. Emerging flux and coronal field is determined as left-handed (LH) or right-handed (RH) as outlined in Sect. \ref{sec:method_chirality}. Estimated heights of the flux ropes that formed in each active region are produced by the method described in Sect. \ref{sec:method_height}.\\
 		\tablefoottext{a}In the case of March 2012, the first three rows represent the orbiting given by the fragment-tracking method. Observations suggest that the full extent of the orbiting from 8--14 March is not detected. \tablefoottext{b}{The fourth row for March 2012 contains orbit angles estimated by-eye.}
 		 }
		\label{tbl:rotation}
\end{table*}

Between 5 March and 7 March, the tracked fragments orbited anticlockwise by 74$\degr$ (an average of $37\degr$ per day), and from 7 March until midday on 8 March, another pair orbited anticlockwise by 36$\degr$ (an average of $\approx29\degr$ per day).
During these times, three X-flares occurred that were associated with halo CMEs: one on 5 March and two in quick succession on 7 March.
The flux ropes that erupted on 7 March were not rooted in the leading, coalescing sunspot and formed low-down in the atmosphere as a result of shearing motions elsewhere in the active region \citep{chintzoglou2015ApJ}. For more on the 7 March eruptions, see \citet{wang2014restructuring}, \citet{syntelis2016spectroscopic}, and \citet{baker2019ifip}.
The total magnetic flux of the active region decayed from 6 March until 10 March, and by 9 March, all of the emerging fragments had coalesced together forming one large sunspot umbra.

On 10 March, the magnetic flux content of the active region began to increase once again, albeit this time without any new orbiting fragments.
Little orbital motion of the previously-merged fragments was measured from 9--15 March (an average of 3$\degr$ per day in the anticlockwise direction).
This measured motion appears to be noisy, consisting of small values $\lesssim 1\degr$ per hour that vary between clockwise and anticlockwise directions (Fig. \ref{fig:rot_goes_flux_201203}).
However, when examining the merged sunspot by eye, there appear to still be two distinct fragments moving closely around each other that are not resolved by the fragment-tracking method (fragments B and C in panels e and f of Fig. \ref{fig:rot_2012march}).
The magnetic flux fragment that arrived at the south of the leading sunspot at 12:00 UT on 8 March (fragment C) orbited almost 180$\degr$ anticlockwise up to the north of the sunspot by 12:00 UT on 11 March -- an average rotation of 60$\degr$ per day.
The CME studied in this section that occurred on 13 March 2012 followed this second phase of flux emergence.

\subsection{13 \& 14 June 2012}

\subsubsection{Coronal evolution} \label{sec:jun2012coronal}

 \begin{figure}
 \centering{}
 \resizebox{\hsize}{!}{\includegraphics{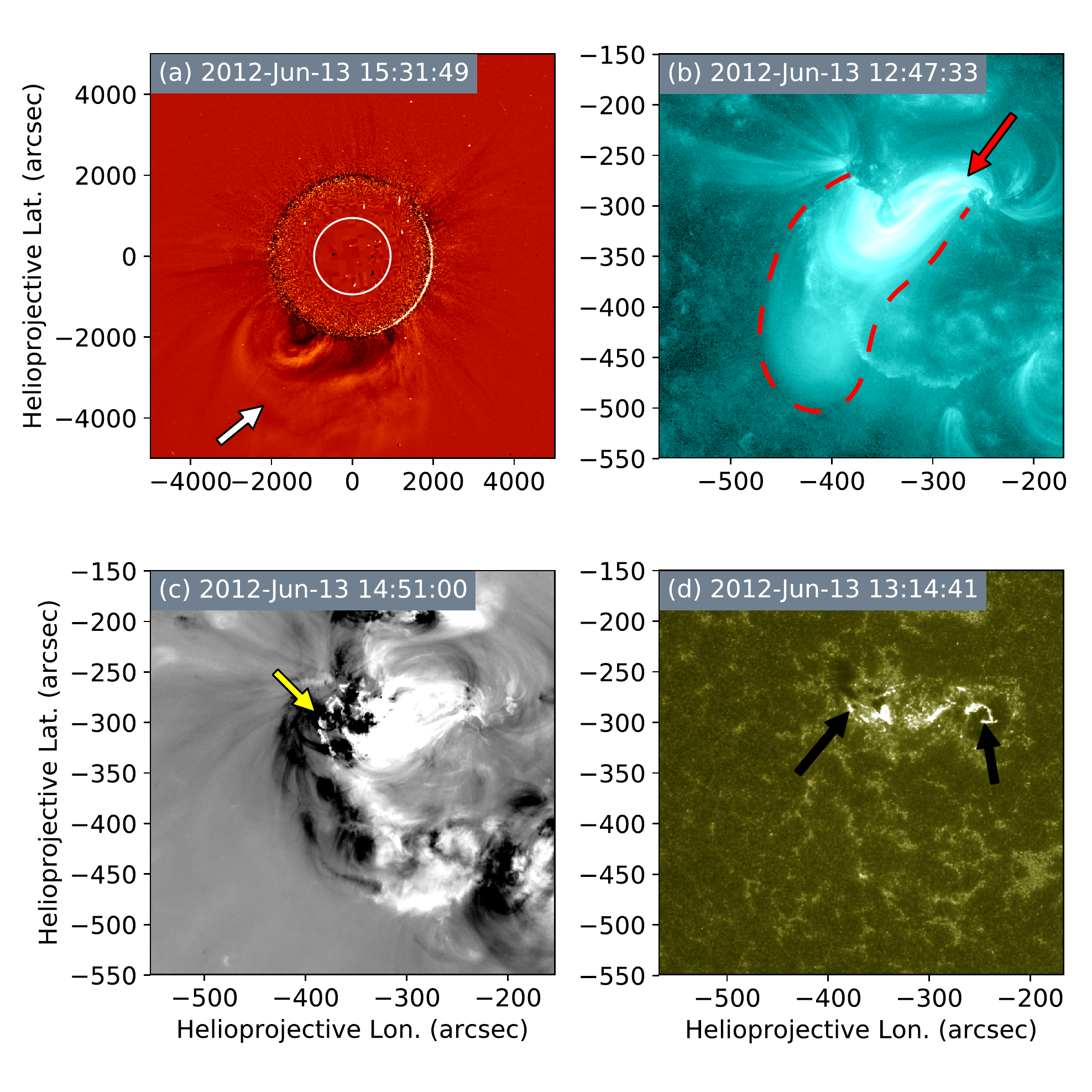}}
 \caption{Observations of the 13 June 2012 eruption. (a) White-light CME (indicated by the white arrow) observed by LASCO C2. (b) A plasmoid (highlighted by a red dashed line) brightens above a flaring arcade (indicated by a red arrow) in the 131 \AA{} channel of AIA. (c) An EUV dimming (marked by a yellow arrow) seen in base difference 211 \AA{} images. (d) Hooked flare ribbons (indicated by two black arrows) seen in the 1600 \AA{} AIA channel.}
 \label{fig:euv_2012june13} 
 \end{figure} 
 
 \begin{figure} 
 \centering{}
 \resizebox{\hsize}{!}{\includegraphics{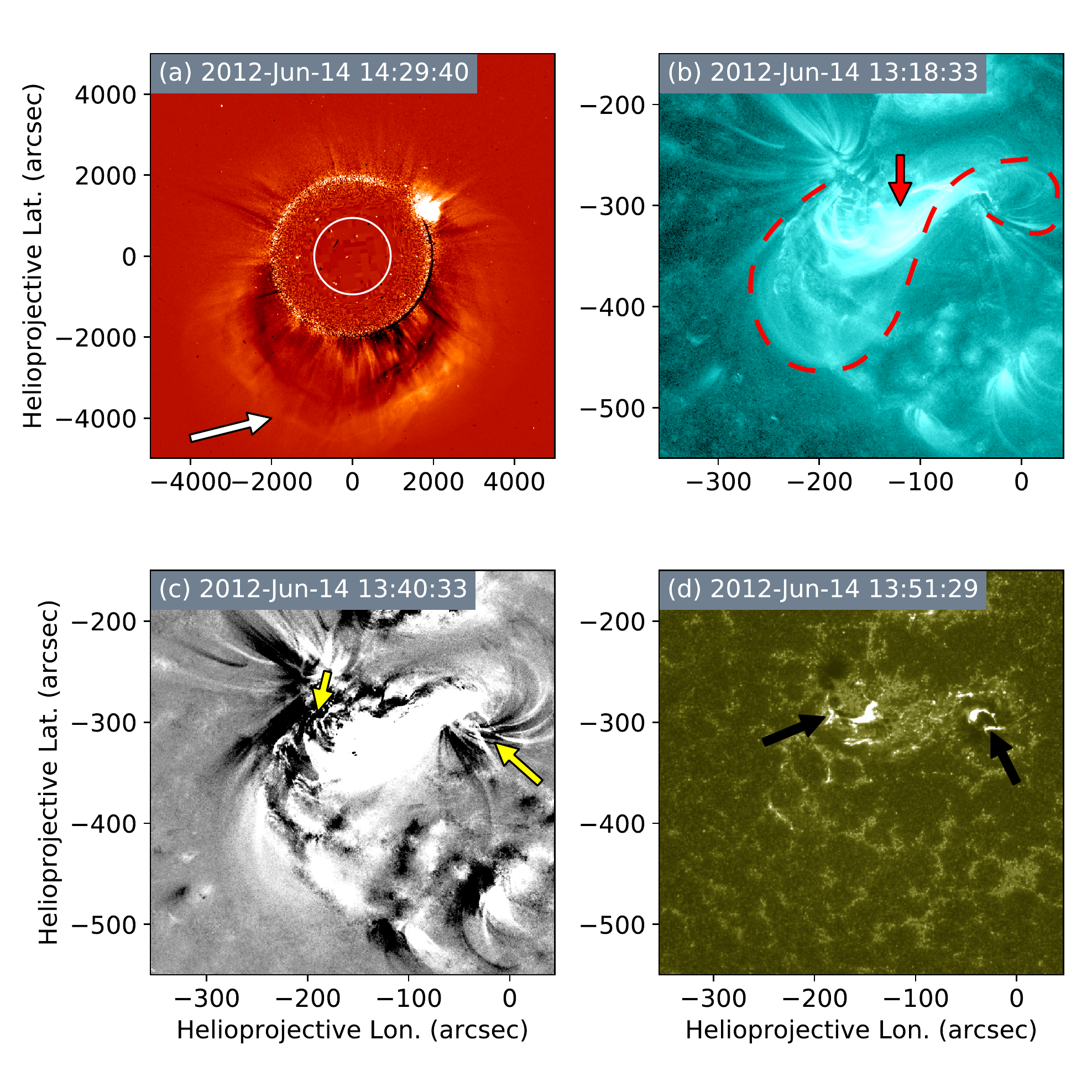}}
 \caption{Observations of the 14 June 2012 eruption. (a) White-light CME (indicated by the white arrow) observed by LASCO C2. (b) A sigmoid (highlighted by the dashed red line) brightens above a flaring arcade (indicated by the red arrow) in the 131 \AA{} channel of AIA. (c) Twin EUV dimmings (marked by two yellow arrows) seen in base difference 131 \AA{} AIA images. (d) Hooked flare ribbons (indicated by two black arrows) seen in the 1600 \AA{} channel of AIA.}
 \label{fig:euv_2012june14} 
 \end{figure}

On 13 June 2012 at 13:00 UT a CME erupted from NOAA AR 11504 in association with an M1.2 GOES class flare. The CME was seen in white-light by LASCO C2 at 14:36 UT (Fig. \ref{fig:euv_2012june13}a). A hot EUV plasma emission feature (plasmoid) started to develop around 11:30 UT and it then rose above the flare arcade (that developed into the arcade of the M1.2 flare) during the 2 hours leading up to its eruption as the CME (Fig. \ref{fig:euv_2012june13}b). 
At 12:14 UT, the underside of the slowly-rising plasmoid is estimated to be at a height of 173 $\pm$ 26 Mm above the centre of the active region's photospheric PIL ($h_{x} = 193 \pm 32\ \textrm{Mm}$, $h_{y} = 153 \pm 21\ \textrm{Mm}$).
An EUV dimming and hooked flare ribbons that formed in the east and west parts of the active region during the eruption infer the footpoint locations of the flux rope (Fig. \ref{fig:euv_2012june13}c and \ref{fig:euv_2012june13}d). 

The previous CME from NOAA AR 11504 occurred on 10 June, seen from 07:30 UT in LASCO C2 images as a faint circular blob. 
An associated M1.3 flare began in the active region at $\approx$06:39 UT on 10 June.
Between the CMEs on 10 June and 13 June, there were 11 flares or brightenings in NOAA AR 11504.
All were fairly weak ($\leq$C2.7 GOES class) and they illuminated loops that spanned the full width of active region as did the 13 June CME and flare.
During seven of these flares, the 1600 \AA{} channel of AIA showed homologous flare ribbons in similar locations to those seen at the onset of the CME on 13 June.
These flares occurred on 11 June at 17:20 UT (no GOES class assigned), on 12 June: a C2.0 flare at 06:38 UT, a C1.6 at 09:00 UT, a C.1.1 at 12:15 UT, and on 13 June: a C2.2 flare at 03:15 UT, a C1.3 at 05:40 UT, and a C2.7 at 09:13 UT.
We suggest these seven homologous confined flares were related to reconnection that built the pre-eruptive flux rope, with the first occurring at $\approx$17:20 UT on 11 June ($\approx$42 hours before the CME on 13 June).

NOAA AR 11504 also produced a CME on 14 June in association with an M1.9 GOES class flare.
As described in Paper I, this CME was seen in LASCO C2 on 14 June 2012 at 14:12 UT (Fig. \ref{fig:euv_2012june14}a).
A sigmoid brightened in the active region above the arcade (that became the loops of the M1.9 class flare) at least 2 hours before erupting (Fig. \ref{fig:euv_2012june14}b).
The middle of the sigmoid is estimated to be at an altitude of 89 $\pm$ 30 Mm above the photosphere at 12:24 UT ($h_{x} = 85 \pm 37\ \textrm{Mm}$, $h_{y} = 94 \pm 22\ \textrm{Mm}$).
For comparison, the underside of the extrapolated flux rope in Paper II was 45 Mm above the photosphere at the same time, and its top was at 150 Mm.
Whilst the estimated sigmoid height is greater than that of the modelled flux rope's underside (where the sigmoid is expected to be), it is lower than the highest point, suggesting that the method used in this study can provide estimates of flux rope height that are at least of comparable orders of magnitude.
EUV dimmings and hooked flare ribbons show the footpoint locations of the flux rope during the eruption (Fig. \ref{fig:euv_2012june14}c, and \ref{fig:euv_2012june14}d). The observations suggest that an HFT flux rope formed in the active region, which is further supported by spectroscopic measurements of coronal plasma and an extrapolated magnetic field model (Papers I and II).

There were six confined brightening events (flares and EUV flux rope signatures) that spanned the active region PIL between the CMEs on 13 and 14 June, with the flares ranging from GOES class C1.7--C5.0 (on 13 June at 19:17 UT and on 14 June at 03:10 UT, 04:44 UT, 07:31 UT, 10:45 UT, and 11:05 UT).
In each case, flare ribbons were observed in the AIA 1600 Å channel that resemble those seen during the 14 June CME.
Therefore, we infer that these flares evidence the formation of the flux rope that erupted on 14 June throughout an 18-hour period.

\subsubsection{Photospheric evolution}

 \begin{figure} 
 \centering{}
 \resizebox{\hsize}{!}{\includegraphics{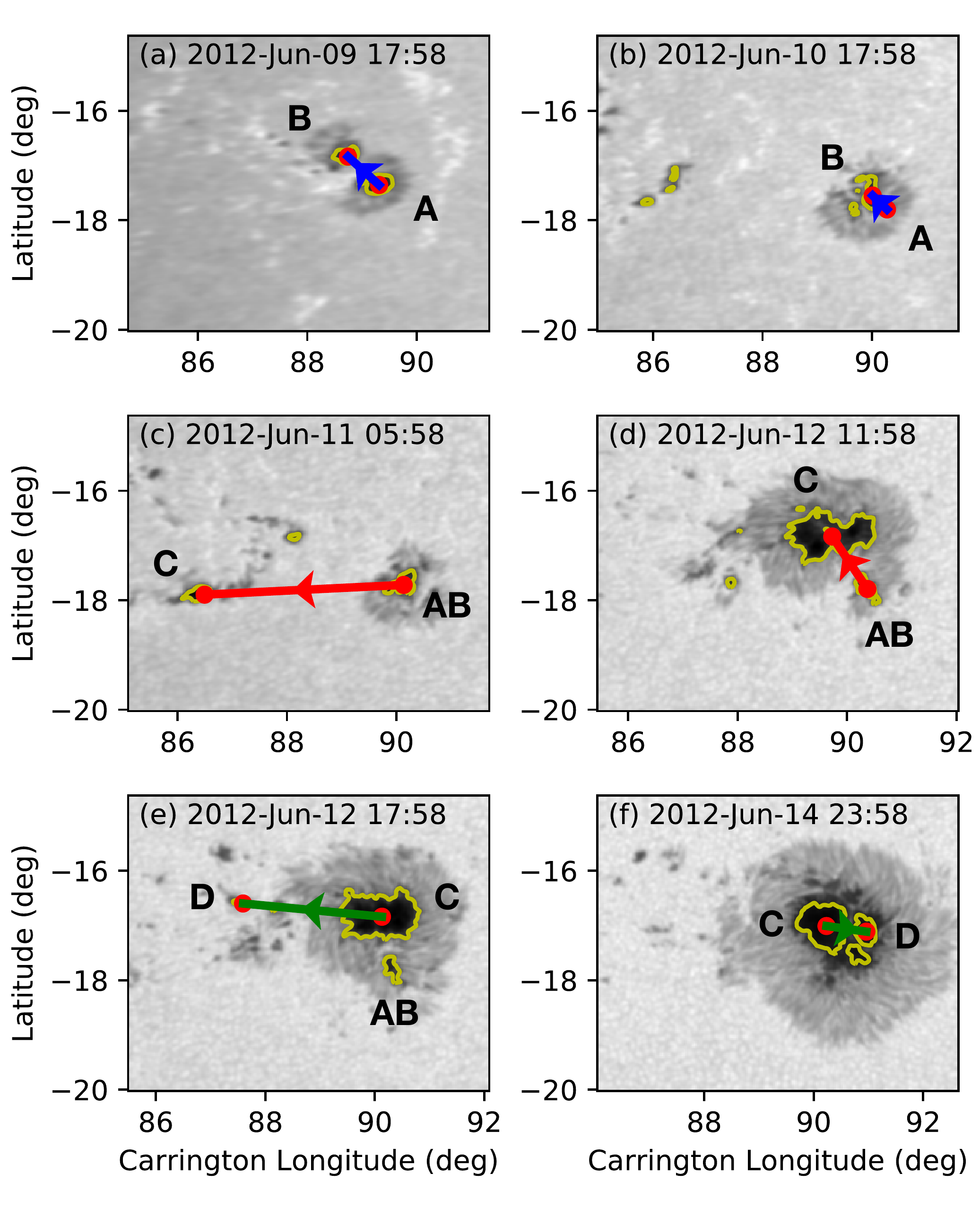}}
 \caption{Clockwise motion of newly-emerged flux around the pre-existing positive (leading) sunspot in NOAA AR 11504. Tracked fragments of flux are labelled A--D. In each panel, vectors are drawn to connect the flux-weighted centroids of two orbiting fragments. This figure is available as an online movie.}
 \label{fig:rot_2012june}
 \end{figure}

 \begin{figure} 
 \centering{}
 \resizebox{\hsize}{!}{\includegraphics{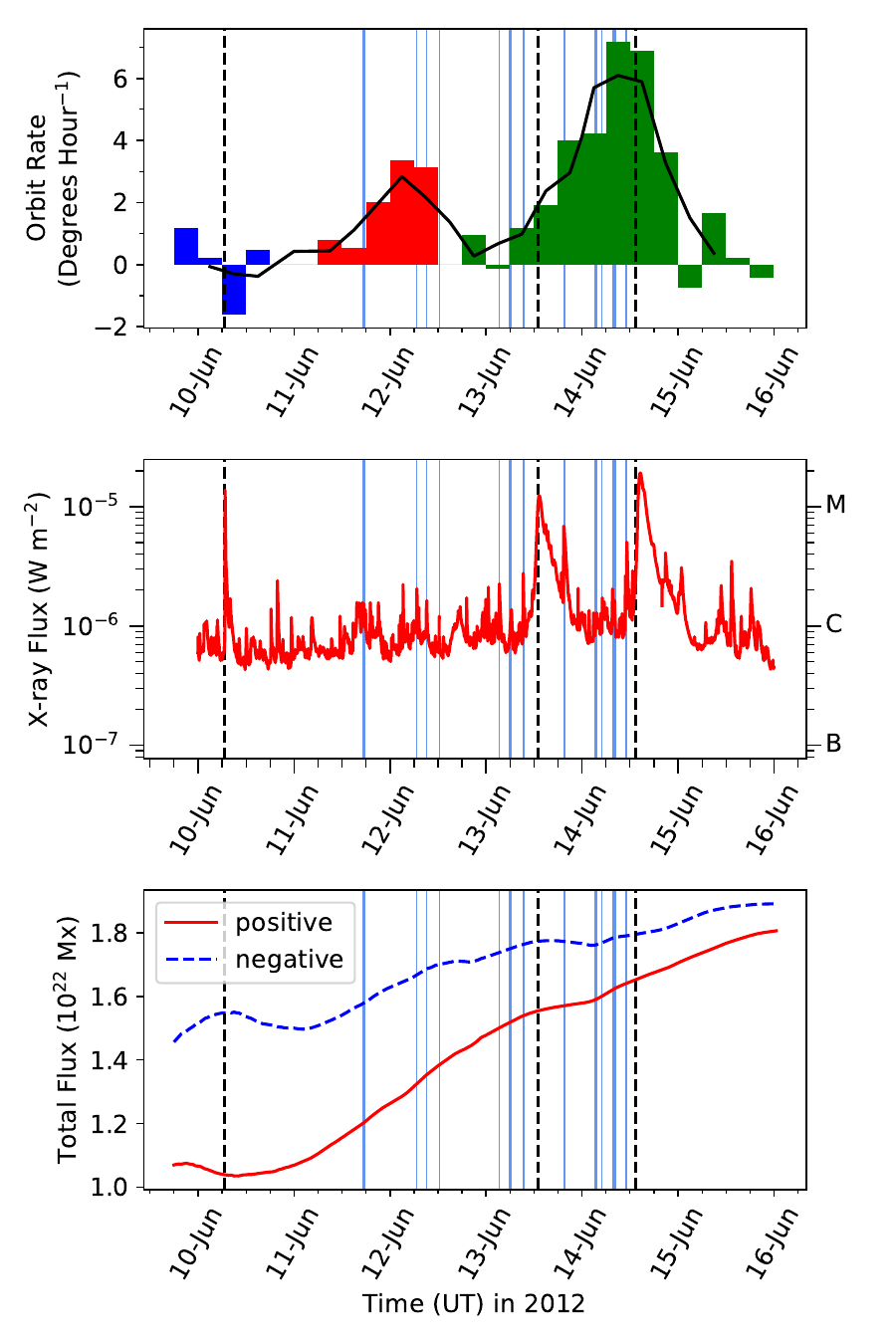}}
 \caption{Top: Measured orbital motion of chosen magnetic flux fragments in 6-hour intervals. The colours correspond to different choices of fragments (see Fig. \ref{fig:rot_2012june}). The black curve represents the orbit angles smoothed with a three-point moving average. Middle: Full-disc integrated GOES soft X-ray lightcurve. Bottom: The evolution of magnetic flux in NOAA AR 11504, made using the radial magnetic field component, $B_{\mathrm{r}}$, of the HMI SHARP data series and smoothed with a 24-hour moving average. Vertical dashed lines indicate the times of the CME onsets on 10, 13, and 14 June 2012, and the thin, light-blue vertical bars show the timings of confined flares associated with flux rope formation described in Sect. \ref{sec:jun2012coronal}.}
 \label{fig:rot_goes_flux_201206}
 \end{figure}

NOAA AR 11504 was close to the centre of the solar disc when it erupted on 13 and 14 June 2012.
This gives an excellent viewpoint from SDO for examining the corona and photosphere with minimal projection effects. As detailed in Paper I, the active region contained two existing sunspots as it rotated on to the solar disc on 8 June, and flux emergence was clearly taking place by 9 June.
The emerging flux showed right-handed magnetic tongues, indicating that the emerging flux tube had right-handed twist (positive chirality). 
The magnetic flux emergence occurred in `episodes', with a major episode that began on 11 June and continued until 15 June in which distinct fragments of magnetic flux emerged one after the other and moved towards and clockwise around the pre-existing sunspots of the same polarities (described in detail in Section 3.2 of Paper I).

The strongest orbital motion was observed around the leading sunspot (see Fig. \ref{fig:rot_2012june}), and in this section we refer to certain fragments of magnetic flux around this sunspot with letters A--D, as in Fig. \ref{fig:rot_2012june}.
Little-to-no significant orbital motion was measured on 9 and 10 June ($2\degr$ in 24 hours between the pre-existing fragment A and the recently-emerged B).
Between 11 June 06:00 UT and 12 June 12:00 UT, a fragment of emerged positive magnetic flux (fragment C) moved clockwise around the pre-existing positive sunspot (the now-merged fragments A and B) by $\approx 60 \degr$ (an average orbital motion of $48 \degr$ per day). 
This emerging fragment (C) grew to become the dominant umbra in the leading sunspot, and between 12 June 18:00 UT and 15 June 18:00 UT, another fragment (D) travelled $\approx 186 \degr$ around fragment C (an average orbital motion of $\approx 62 \degr$ per day). 
These quantified orbits are given in Table \ref{tbl:rotation} and also represented in the top panel of Fig. \ref{fig:rot_goes_flux_201206}.

Fig. \ref{fig:rot_goes_flux_201206} shows that the two eruptive M-class flares from the active region occurred during the second period of strongest orbiting. 
Between the CMEs on 10 and 13 June, a net clockwise orbiting of $74\degr$ was observed, and between the CMEs on 13 and 14 June, $145\degr$ of clockwise orbiting occurred. 
This correlation between stronger orbiting and shorter formation time between eruptions supports the hypothesis that the orbital motions bring coronal loops together and facilitate the magnetic reconnection that builds the flux ropes.

\subsection{8 October 2012}

\subsubsection{Coronal evolution} 
\label{sec:oct2012coronal}

 \begin{figure} 
 \centering{}
 \resizebox{\hsize}{!}{\includegraphics{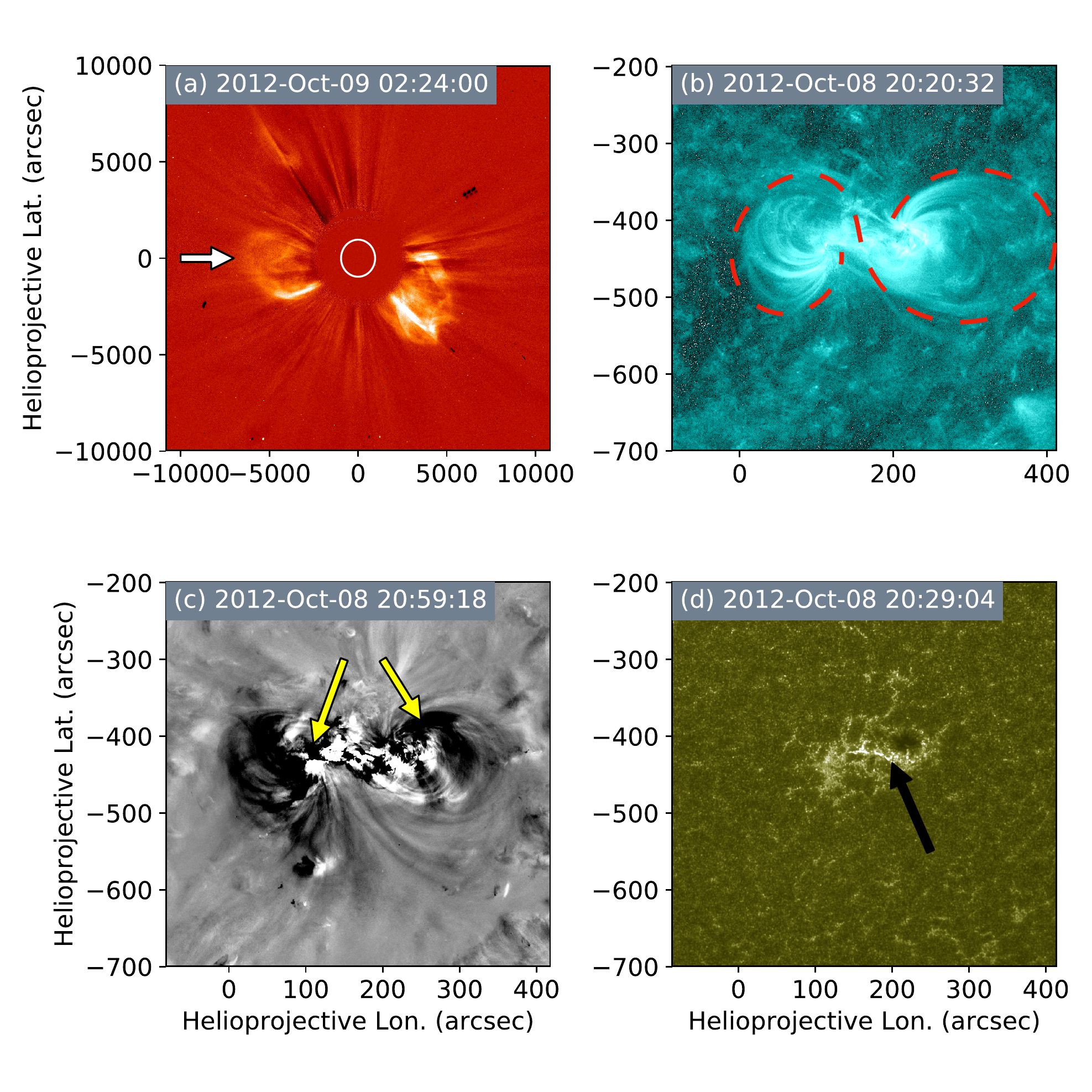}}
 \caption{Observations of the 8 October 2012 eruption. (a) White-light CME (indicated by the white arrow) observed by STEREO COR2-A. (b) An inferred sigmoid (indicated by the dashed red line) in the 131 \AA{} channel of AIA. (c) Twin EUV dimmings (marked by the two yellow arrows) seen in base difference 193 \AA{} images. (d) One hooked flare ribbon (indicated by the black arrow) seen in the 1600 \AA{} AIA channel.}
 \label{fig:euv_2012october8} 
 \end{figure}

A CME erupted from NOAA AR 11585 at $\approx$20:30 UT on 8 October 2012.
The eruption was linked to a slow white-light CME seen in STEREO-A coronagraphs: COR1 from 21:15 UT, and COR2 from 23:09 UT (Fig. \ref{fig:euv_2012october8}a).
A faint arcade brightened in the active region as the eruption proceeded, but it was not designated a GOES flare class.
Loops that extended from the east and west of the active region appeared strongly curved in the hours before the eruption, and from 19:30 UT, both of the curved ends began to expand, suggesting an expanding sigmoid in the active region (Fig. \ref{fig:euv_2012october8}b). The interpretation of a continuous sigmoid in the active region is supported by X-ray observations\footnote{\url{https://darts.isas.jaxa.jp/solar/hinode/query.php?A25=ql&inst[]=xr&plot=no&tmSY=2012&tmSM=10&tmSD=07&tmSh=11&tmSm=00&tmEY=2012&tmEM=10&tmED=07&tmEh=23&tmEm=40&fgA=basic&fgB[]=0&fgE=0&spA=basic&spB[]=0&xrA=basic&xrB[]=0&xrD[]=1&eiA=basic&eiB[]=0&max=1000&xrpd[]=0&idx=5}} taken on 7 October 2012 by the \textit{Hinode} \textit{X-Ray Telescope} \citep{golub2007xrt}.
Furthermore, hot, rising plasma emission was seen near the centre of the active region from 20:19 UT, suggesting that a flux rope had formed in the active region. 
The hot feature was at an estimated altitude of $98 \pm 25\ \textrm{Mm}$ above the centre of the photospheric PIL at 20:19 UT ($h_{x} = 112 \pm 39\ \textrm{Mm}$, $h_{y} = 85 \pm 12\ \textrm{Mm}$).
During the eruption, twin EUV dimmings and a hooked flare ribbon developed in the active region, indicating that the erupting flux rope had its footpoints in the east and west sides of the region (Fig. \ref{fig:euv_2012october8}c and \ref{fig:euv_2012october8}d).

No previous CMEs were observed to have originated from NOAA AR 11585 during its passage across the solar disc. 
Therefore, the entire disc-passage of the region from its appearance on 1 October until the CME on 8 October was analysed to search for flux rope formation signatures.
Five confined flares occurred in the centre of NOAA AR 11585 in the same place as the subsequent eruption arcade, ranging from GOES class B3.2--B5.3 (three on 6 October at 05:23 UT, 07:06 UT, and 11:19 UT, and two on 7 October at 03:16 UT and 12:22 UT), all with similar flare ribbons that resemble the ones seen during the 8 October CME.
This suggests the confined flares were involved in the formation of the flux rope that erupted on 8 October, and therefore they indicate the times at which the flux rope formed. In this way, formation started with a confined flare at 05:23 UT on 6 October, $\approx$63 hours before the 8 October CME.

\subsubsection{Photospheric evolution}

 \begin{figure} 
 \centering{}
 \resizebox{\hsize}{!}{\includegraphics{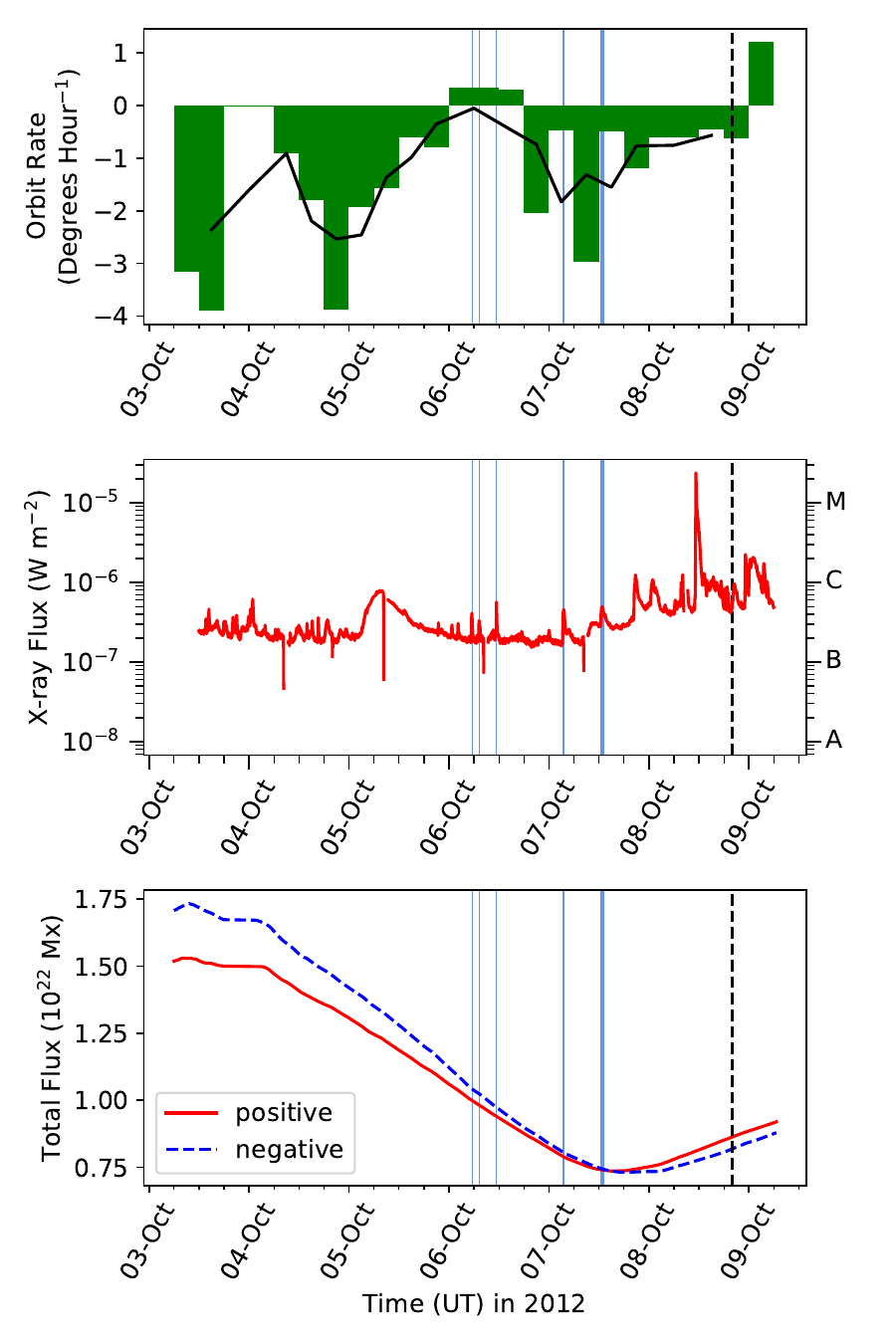}}
 \caption{Top: Measured orbital rotation of a chosen magnetic flux fragment around another in 6-hour intervals (see Fig. \ref{fig:rot_2012october}). The black curve represents the orbit angles smoothed with a three-point moving average. Middle: Full-disc integrated GOES soft X-ray lightcurve. Bottom: The evolution of magnetic flux in NOAA AR 11585, made using the radial magnetic field component, $B_{\mathrm{r}}$, of the HMI SHARP data series and smoothed with a 24-hour moving average. The vertical dashed line indicates the time of the CME onset on 8 October 2012, and the thin, light-blue vertical bars show the timings of confined flares associated with flux rope formation described in Sect. \ref{sec:oct2012coronal}.}
 \label{fig:rot_goes_flux_201210}
 \end{figure}
 
 \begin{figure} 
 \centering{}
 \resizebox{\hsize}{!}{\includegraphics{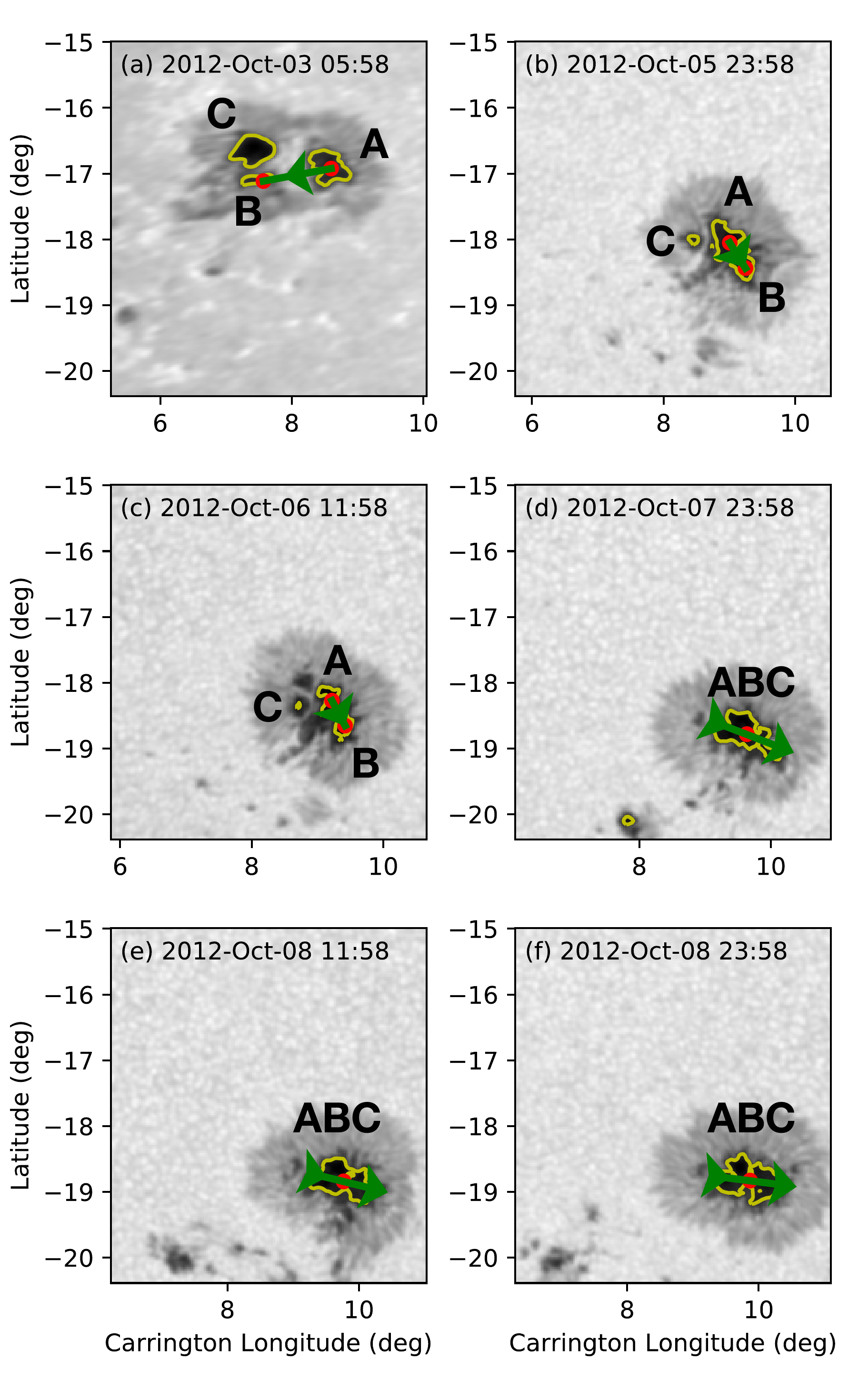}}
 \caption{Anticlockwise motion of newly-emerged flux around the pre-existing positive (leading) sunspot in NOAA AR 11585. Tracked fragments of flux are labelled A--C. In each image, vectors are drawn either to connect the flux-weighted centroids of two orbiting fragments (panels a-c) or to best-fit the major axis of merged fragments (panels d-f). This figure is available as an online movie.}
 \label{fig:rot_2012october}
 \end{figure}

The leading sunspot of NOAA AR 11585 was forming as it rotated on to the solar disc and a small trailing (negative) spot briefly coalesced on 3 October before dispersing.
The region exhibited left-handed magnetic tongues, implying the emerging flux had left-handed twist (negative chirality).
Flux emergence had ceased by 4 October, and the region began to decay (see Fig. \ref{fig:rot_goes_flux_201210}).

The positive sunspot was comprised of three distinctly-emerged umbral fragments within a single penumbra that orbited around each other during the week before the eruption (labelled A--C in Fig. \ref{fig:rot_2012october}).
The two fragments that showed the strongest orbiting relative to each other were chosen (A and B).
It is worth emphasising that the majority of the orbiting motions observed in this event occurred when the active region was decaying, in other words, after flux emergence had finished (or at least after the horizontal part of the emerging flux tube had crossed the photosphere).

Between 3 October 06:00 UT and 6 October 00:00 UT (2.75 days), the chosen fragments orbited by $111\degr$ (an average of $\approx40\degr$ per day). 
Between 6 October 12:00 UT and 8 October 00:00 UT (1.5 days), the chosen fragment orbited around the other by $41\degr$ (an average of $\approx27\degr$ per day) and merged into one umbra (labelled ABC in Fig. \ref{fig:rot_2012october}).
Between 8 October 12:00 UT and 9 October 00:00 UT (0.5 days), the merged umbra rotated by $7\degr$ (an average of $14\degr$ per day). 
These motions are quantified in Table \ref{tbl:rotation} and visualised in Fig. \ref{fig:rot_goes_flux_201210} along with the GOES X-ray activity of the Sun during these times.

From 3--7 October, the active region was decaying, but it entered a new phase of emergence from 7 October until past the time of eruption.
$159\degr$ of anticlockwise orbiting was measured from 3--9 October -- an average of $\approx 26.5\degr$ per day.
The orbiting of fragments in this event was weaker than in some others, but occurred continuously over a long time.
This could explain the lack of flux rope ejections from this active region, with weaker motions taking a longer time to build a flux rope.

\subsection{14 July 2017}

\subsubsection{Coronal evolution} \label{sec:jul2017coronal}

 \begin{figure} 
 \centering{}
 \resizebox{\hsize}{!}{\includegraphics{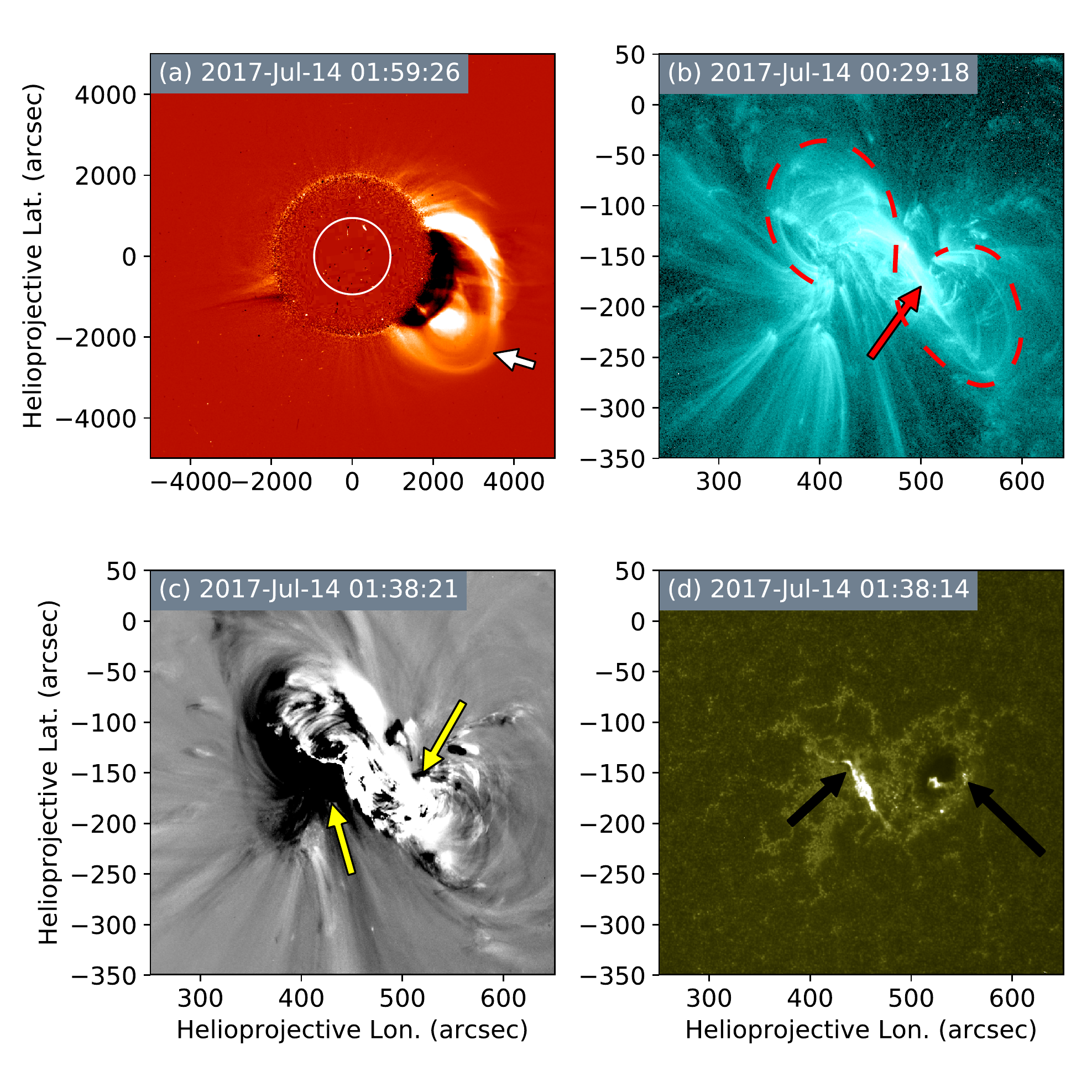}}
 \caption{Observations of the 14 July 2017 eruption. (a) White-light CME (indicated by the white arrow) observed by LASCO C2. (b) Curved loops (indicated by the dashed red lines) that may belong to a sigmoid above the observed flare arcade (indicated by the red arrow) in the 131 \AA{} channel of AIA. (c) Twin EUV dimmings (marked by the two yellow arrows) seen in base difference 211 \AA{} images. (d) Flare ribbons (indicated by two black arrows) seen in the 1600 \AA{} channel of AIA.}
 \label{fig:euv_2017july14}
 \end{figure} 

On 14 July 2017, a CME erupted from NOAA AR 12665 at $\sim$01:00 UT, accompanied by an M2.4 flare.
The corresponding white-light CME was first seen by LASCO C2 at 01:36 UT (Fig. \ref{fig:euv_2017july14}a).
The active region appeared faintly sigmoidal several hours before the eruption in the 131 \AA{} channel of AIA (Fig. \ref{fig:euv_2017july14}b).
A relatively faint flare arcade began to appear beneath the sigmoid from $\sim$00:00 UT, suggesting an HFT flux rope was present in the active region before the CME.
The sigmoid began to expand and erupt at 00:30 UT, and the flare arcade brightened and grew in to a cusp shape. 
At 00:30 UT, the peak of the flare arcade cusp was at an estimated $71 \pm 20\ \textrm{Mm}$ above the photosphere ($h_{x} = 59 \pm 8\ \textrm{Mm}$, $h_{y} = 82 \pm 32\ \textrm{Mm}$). 
During the eruption, observed twin EUV dimmings and flare ribbons imply that the erupting flux rope had footpoints in the east and west parts of the active region (Fig. \ref{fig:euv_2017july14}c and \ref{fig:euv_2017july14}d).

There were no CMEs observed from NOAA AR 12665 before the one on 14 July, so we look back over the full disc-passage of the region for signs of flux rope formation.
Shortly after a confined M1.3 flare on 9 July, there were two confined C-class flares: a C1.6 flare from 06:15--06:40 UT and a C1.2 flare from 07:28--07:49 UT. Around the times of these C-flares (between 05:35--06:31 UT and 07:51--08:47 UT), faint emission signatures were seen towards the eastern edge of the active region in the 131 \AA{} and 94 \AA{} AIA channels only (suggesting plasma at $\textrm{T} \sim$ 10 MK). We interpret these signatures as part of a hot flux rope, brightening in association with the confined flares. 
Since there were no successful eruptions from 9 July until the CME on 14 July, the hot flux rope evidenced on 9 July may be the same one that erupted on 14 July. This would suggest flux rope formation had begun 115 hours before the CME, although it is difficult to support this by comparing flare ribbons between 9 and 14 July due to the dramatic photospheric evolution that occurred between those times (the strong orbiting described in Section \ref{sec:jul2017photo}).
After the two C-flares described above, eleven additional confined B- or C-class flares occurred on 9 July, followed by a confined C1.5 flare on 10 July. 
No hot flux rope signatures were observed in association with these flares, so we do not link them to flux rope formation.
The earliest flare ribbons that appeared homologous to the eruptive flare were seen during a confined B4.9 flare at 02:00 UT on 11 July, with one bright globular ribbon visible near the leading sunspot (and flux rope footpoint) and a second ribbon near the trailing polarity. Two more confined B-flares (at 04:40 UT on 11 July and 05:15 UT on 13 July) also showed similar homologous flare ribbons, suggesting they, too, were involved in forming the flux rope. In summary, there is evidence to suggest that two C-flares on 9 July, two B-flares on 11 July, and a B-flare on 13 July were involved in the formation of the flux rope that erupted on 14 July 2017.

\subsubsection{Photospheric evolution} \label{sec:jul2017photo}

 \begin{figure} 
 \centering{}
 \resizebox{\hsize}{!}{\includegraphics{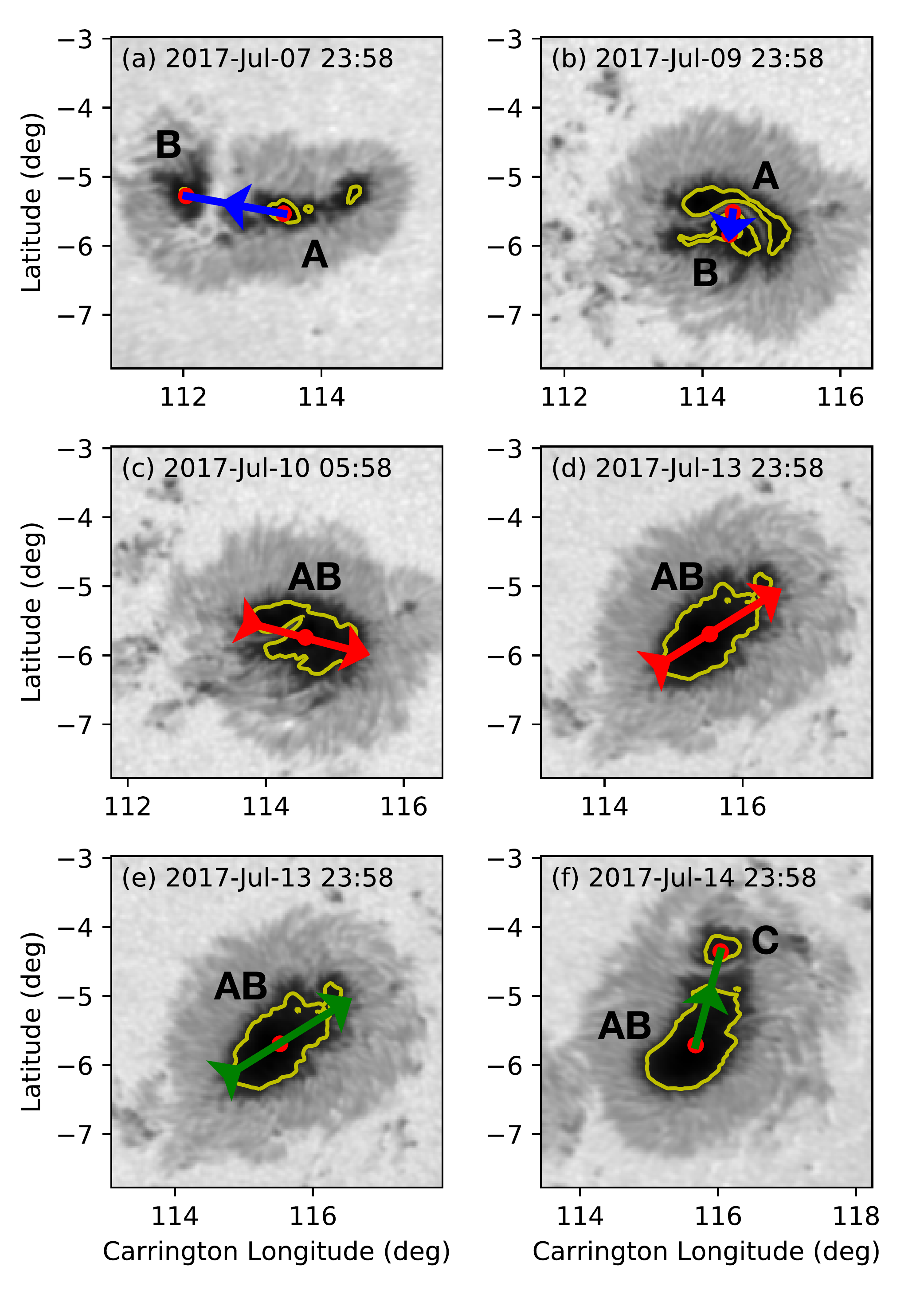}}
 \caption{Anticlockwise motion of newly-emerged flux around the pre-existing positive (leading) sunspot in NOAA AR 12665. Tracked fragments of flux are labelled A--C. In each image, vectors are drawn either to connect the flux-weighted centroids of two orbiting fragments (panels a, b, and f) or to best-fit the major axis of merged fragments (panels c, d, and e). This figure is available as an online movie.}
 \label{fig:rot_2017july}
 \end{figure}

 \begin{figure}
 \centering{}
 \resizebox{\hsize}{!}{\includegraphics{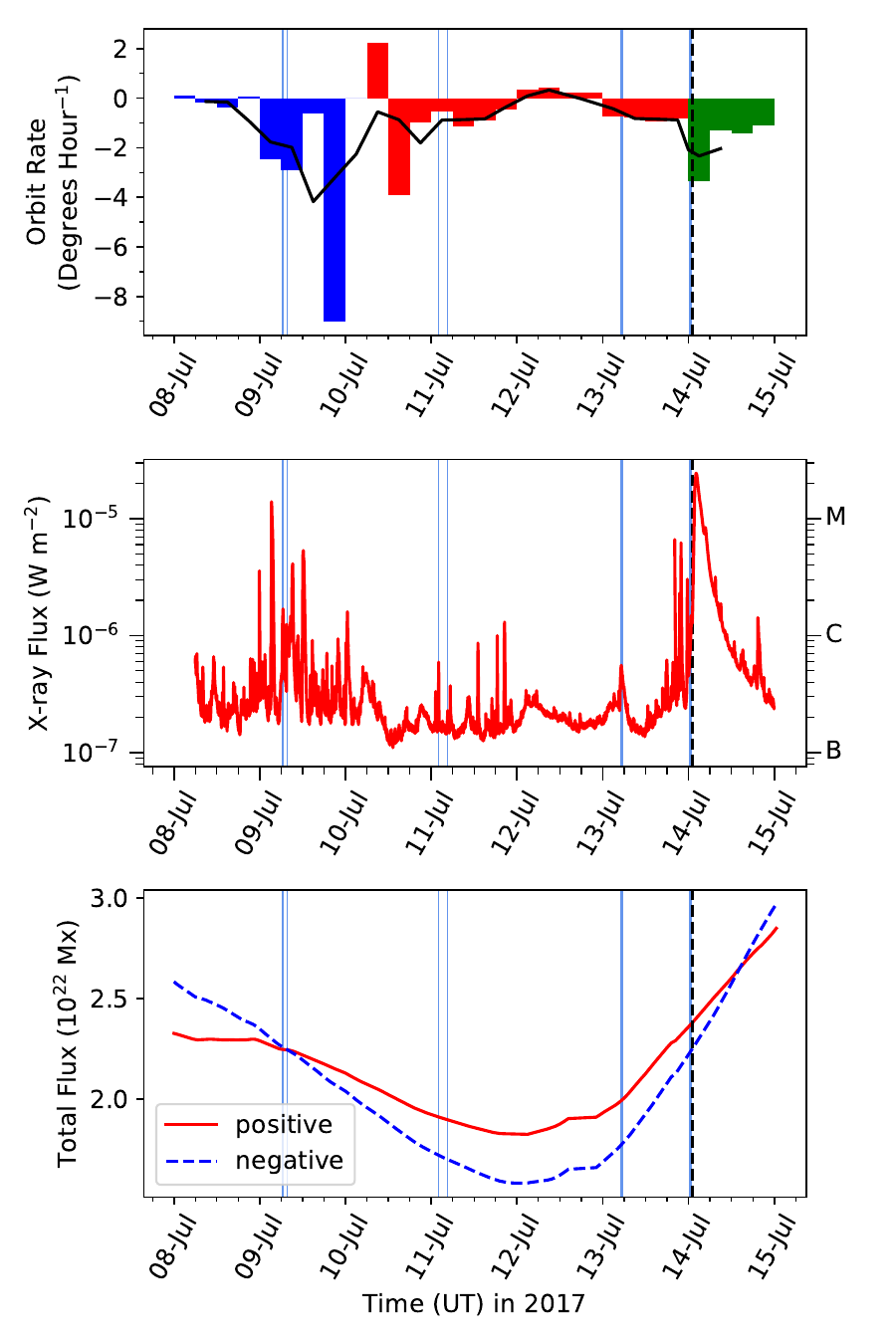}}
 \caption{Top: Measured orbital motion of chosen magnetic flux fragments in 6-hour intervals. The colours correspond to different choices of fragments (see Fig. \ref{fig:rot_2017july}). The black curve represents the orbit angles smoothed with a three-point moving average. Middle: Full-disc integrated GOES soft X-ray lightcurve. Bottom: The evolution of magnetic flux in NOAA AR 12665, made using the radial magnetic field component, $B_{\mathrm{r}}$, of the HMI SHARP data series and smoothed with a 24-hour moving average. The vertical dashed line indicates the time of CME onset on 14 July 2017, and the thin, light-blue vertical bars show the timings of confined flares associated with flux rope formation described in Sect. \ref{sec:jul2017coronal}.}
 \label{fig:rot_goes_flux_201707}
 \end{figure}

As the region rotated on to the disc on 5 July, it featured two small pre-existing sunspots; a leading positive spot and a trailing negative.
Observed magnetic tongues demonstrate that the magnetic flux that emerged in to the active region was left-handed (negative chirality). Some very weak anticlockwise orbiting can be observed around the trailing sunspot, but as with the other events, here we focus on the leading sunspot.

A fragment of magnetic flux emerged on 6 July (fragment B in Fig. \ref{fig:rot_2017july}) and moved towards the leading sunspot before rotating anticlockwise around the pre-existing sunspot umbra (fragment A).
Between 8 July 00:00 UT and 10 July 00:00 UT (2 days), the emerging fragment orbited around the other by $92\degr$ (an average of $46\degr$ per day).
The strongest orbiting during this period was seen on the evening of 9 July, where the fragments sheared past each other from 18:00--00:00 UT for a rotation of $54\degr$.
Between 10 July 06:00 UT and 14 July 00:00 UT (3.75 days), fragments A and B merged together, and an orbit of $47\degr$ is inferred from the continued rotation of the sunspot (an average of $12.5\degr$ per day).
Between 14 July 00:00 UT and 15 July 00:00 UT (1 day), a fragment broke away from the merged umbra (labelled fragment C) and continued to orbit the sunspot by $43\degr$.
These values are collected in Table \ref{tbl:rotation}, and presented in Fig. \ref{fig:rot_goes_flux_201707} with the corresponding X-ray activity measured by GOES.
 
From 8--14 July, $138\degr$ of anticlockwise orbiting was measured -- an average of $\approx 20\degr$ per day.
The average daily orbiting in this active region is weaker than in some events, but since there were no observed prior CMEs from the region, and the flux rope may have been forming since as early as 9 July, the flux rope may have formed gradually over several days.
The main emerging sunspot fragment encountered the pre-existing sunspot on 8 July and showed strong shearing motion clockwise around it throughout 9 July ($90\degr$ in 24 hours).
Most of the strong flares from the active region occurred during this time, with an M2.4 flare, 4 C-flares, and 16 B-flares.

\section{Discussion} \label{sec:sci3_discussion}

Across Papers I and II, a methodology is detailed by which a combination of observations and modelling enables the identification of hot flux ropes when they are on the solar disc (as opposed to limb detection).
This allowed for the investigation of the formation mechanism of hot flux ropes.

\begin{figure} 
 \centering{}
 \resizebox{\hsize}{!}{\includegraphics{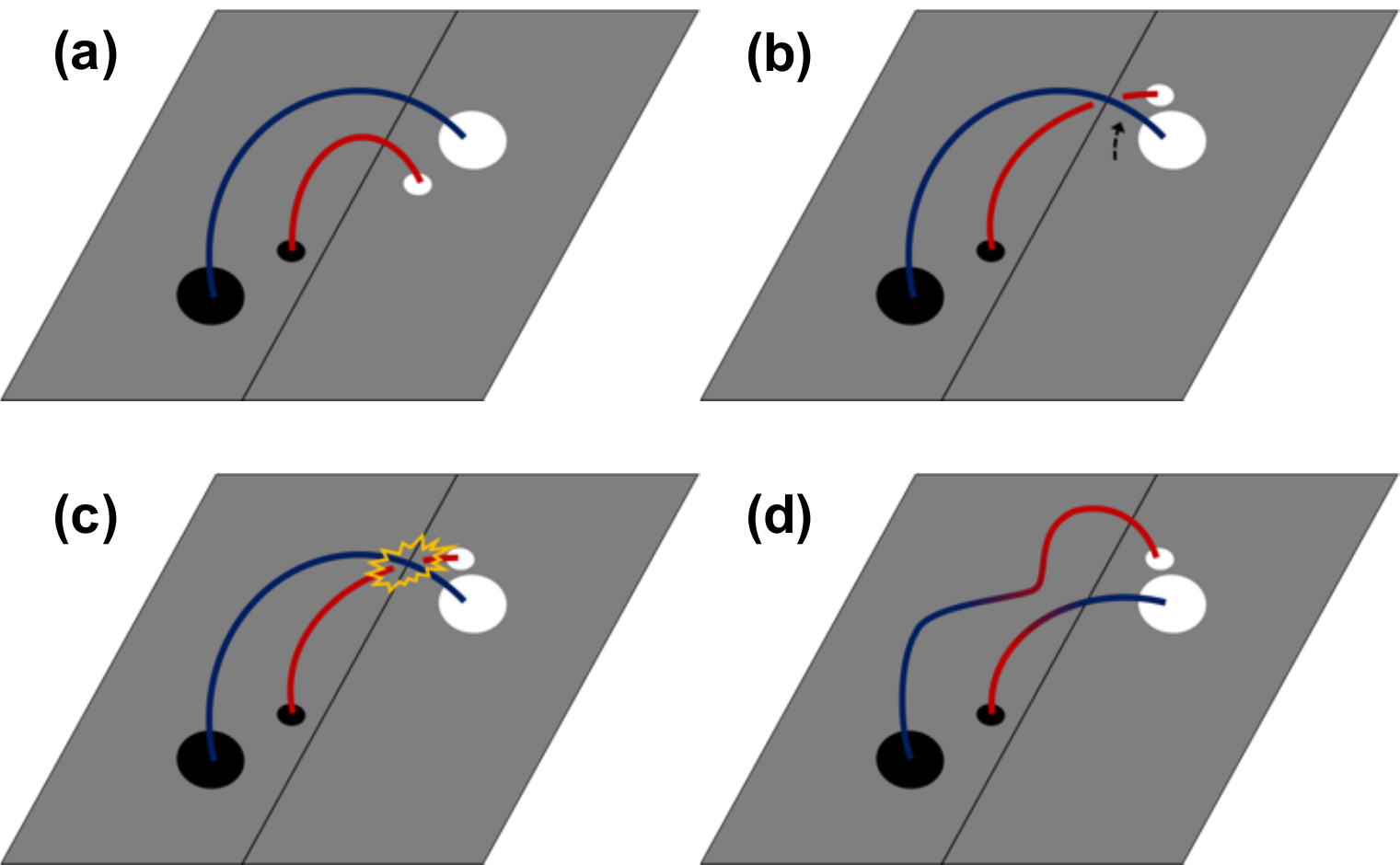}}
 \caption{Orbiting motions of a magnetic flux fragment around a pre-existing sunspot. (a) A magnetic bipole (red) emerges beneath the pre-existing bipole field of sunspots (blue). (b) A fragment of the emerging flux moves towards and `orbits' around the pre-existing sunspot of the same polarity, wrapping field lines around each other. (c) Component magnetic reconnection occurs in the corona. (d) The products of this reconnection are sheared (twisted) field lines in the form of an overlying flux rope and an underlying arcade.}
 \label{fig:orbit_reconnection}
 \end{figure}

This work tests the hypothesis that the photospheric orbiting motions associated with emerging magnetic flux are the trigger for building HFT flux ropes via reconnection in the corona (see Fig \ref{fig:orbit_reconnection}), as was suggested in Papers I and II. 
We have shown that this mechanism is responsible for building five hot flux ropes with the observational signatures of hyperbolic flux tube (HFT) configurations that then erupted as CMEs on 13 March 2012 (from NOAA active region 11429), 13 June 2012 (from NOAA active region 11504), 14 June 2012 (from NOAA active region 11504), 8 October 2012 (from NOAA active region 11585), and 14 July 2017 (from NOAA active region 12665). 
The radial magnetic flux of each active region (between $\sim 3\times10^{21}\ \textrm{Mx}$ and $3\times10^{22}\ \textrm{Mx}$) places them in the large active region category according to \citet[][presented in Table 1 of \citealp{vanDrielGesztelyi2015evolution}]{schrijver2000book}.

% Justification for HFT classification
The classification as an HFT rope comes from the estimated height of the underside of the observed hot flux rope or sigmoid and the presence of lower-lying arcade field as determined from EUV images. As detailed in Table \ref{tbl:rotation}, the heights of the flux ropes that erupted as CME events on 13 March 2012, 13 June 2012, 14 June 2012, 8 October 2012 and 14 July 2017 were estimated to be 56 $\pm$ 5 Mm, 173 $\pm$ 26 Mm, 89 $\pm$ 30 Mm, 98 $\pm$ 25 Mm and 71 $\pm$ 20 Mm above the photosphere, respectively. 
However, the 13 March 2012 estimate may be strongly affected by projection effects, and the height determination in the 13 June 2012 case was made during the slow-rise phase of the structure, potentially leading to an overestimation of the height.
Furthermore, the height of the underside of the flux rope that erupted on 14 July 2017 was determined from the highest point of the flare arcade associated with the event, which is expected to be just beneath the bottom of the flux rope, and the height measurement was also made during the very early stage of the eruption.
The height estimates are of similar order to the height of the flux rope modelled in Paper II, which had its underside and top at 45 Mm and 150 Mm above the photosphere, respectively, and with the observational study of \cite{patsourakos2013direct} who found a hot flux rope that formed at an altitude of 80 Mm during a confined flare, rose slowly, and had an altitude of 138 Mm at the start of the impulsive CME acceleration phase.
The heights place the ropes in the $\beta < 1$ region of the solar atmosphere (see Figure 3 of \citealp{gary2001plasma}), implying that plasma does not contribute to the stability of these HFT flux ropes in the same way that it would if their undersides were line-tied to dense photospheric or chromospheric plasma in a BPS configuration or if they contained dense filament plasma \citep{jenkins2018mass,jenkins2019modeling}.
If perturbed, reconnection could occur at the HFT under these flux ropes, aiding the eruption of the structure.

% Orbiting motion and reconnection as CME trigger
The main finding in this work is that the combination of orbital motions of photospheric magnetic field and magnetic reconnection in the corona is able to act collectively as a CME trigger mechanism, forming pre-eruptive HFT ropes.
In addition to the orbiting motions, all regions produced confined flares that could be termed homologous: the confined flares occurred along the same polarity inversion line, had broadly the same spatial extent as observed in the early stages of the eventual eruptive flares, and featured flare ribbons that resembled those seen later during the CMEs.
This indicates that the formation process of coronal HFT flux ropes in this study is systematically related to magnetic reconnection driven by the photospheric orbiting motions. 

The confined flare observations can therefore be used to probe the timescale over which each flux rope formed; each reconnection event transforming the configuration from sheared arcade to flux rope and feeding more axial flux into the structure. 
Homologous C-class flares before the 13 March 2012 CME (on 12 March at 22:20 UT and 13 March at 06:55 UT) suggest a flux rope was forming in the active region for $\approx$19 hours before it erupted.
In the case of the 13 June 2012 CME, homologous, confined C-flares occurred on 11, 12, and 13 June, suggesting a flux rope formation time of up to 42 hours. Another CME occurred from the same active region on 14 June, with relevant C-class flares spread over an $\approx$18 hour period. Clear signatures of the EUV flux rope in the from of a sigmoid were observed 2 hours before the eruption (see Paper I) indicating its presence as a coherent structure by this time. No other CMEs from the same source region were observed before the one on 8 October 2012. B-class flares were seen in the active region from the morning of 6 October onwards with homologous flare ribbons to those seen during the CME, suggesting a flux rope formation timescale of 63 hours. However, all observational signatures in this active region were weak so there is uncertainty in this proposed timescale.
Finally, faint flux rope signatures were observed in the hottest AIA channels (131 \AA, 94 \AA) during confined C-class flares on 9 July 2017, 115 hours before the CME on 14 July.

In summary, the timescale over which the HFT rope in each active region is likely being built through successive reconnection events varies significantly from 18--115 hours.
The key point here is that several episodes of magnetic reconnection in the corona may be needed to build the HFT structure that becomes eruptive, in an analogous way to the ongoing episodes of magnetic reconnection in the photosphere or chromosphere that build BPS ropes at lower altitudes \citep{van1989formation}.

% Consistency of chirality
The observed photospheric motions were quantified using the method described in Sect \ref{sec:method_orbit}. Across the studied intervals presented in Table \ref{tbl:rotation}, rates of averaged daily orbiting were found between 12--48$\degr$ per day.
The highest rate of orbiting measured between any two subsequent time steps was 54.1$\degr$, measured over a 6-hour interval beginning at 18:00 UT on 9 July 2017 in NOAA active region 12665 (see Figure \ref{fig:rot_goes_flux_201707}). We also note the strong rate of orbiting observed in NOAA active region 11504 between the CMEs on 13 and 14 June 2012. In this $\approx$24 hour period, 103$\degr$ of orbiting was observed (measured from noon on 13 June until noon on 14 June).
This high rate of orbiting may explain how a flux rope was able to form in NOAA AR 11504 so quickly after the previous CME the day before.
However, when considering all of the events together, there are no apparent correlations between formation time and average orbit rate, total orbit angle, or active region flux. Future work could examine a larger sample of active regions to develop a clearer sense of possible relationships between such parameters.

The conjecture that the photospheric orbiting motions are ultimately responsible for the flux rope formation is further tested by comparing the direction of these motions with the handedness of the flux rope.
The observations show that there is consistency between the handedness of the emerging magnetic flux, the direction of the orbiting motion, and the handedness of the flux ropes that form. Active regions with emerging right-handed (left-handed) twist exhibited dominantly clockwise (anticlockwise) orbiting motions in the photosphere, and right-handed (left-handed) flux ropes formed in the corona.
As an aside, we note here that active region NOAA 11429 (March 2012) does not conform to the expected Hale orientation for the northern hemisphere in solar cycle 24, and there is still agreement between emerging twist, orbiting direction, and flux rope handedness in this region. Previous studies have found that sunspot rotation can play a role in building the twist of forming flux ropes \citep{vemareddy2016rotation,yan2018motion}. Similarly, in this study, once orbiting fragments had coalesced, their continuing orbital motions at times resembled the scenario of a rotating sunspot, giving further support to the scenario that twisted magnetic flux ropes can be formed in these active regions. Furthermore, simulations show that rotating sunspots can cause the inflation of overlying magnetic fields in active regions, facilitating the onset of the ideal torus instability as a CME driver \citep{torok2013initiation}.

% Discussion about origin of orbiting motions
The origin of the orbiting motions seen in the photosphere of the active regions under study here is an open and interesting question, which may be related to observations of bodily rotation of sunspots.
\citet{brown2003rotating} found sunspot rotations of $40\degr\textrm{--}60\degr$ over 3 to 5 days, and posed two possible explanations for sunspot rotation: photospheric flows that lead to localised proper motions, or flux emergence. 
Extending these explanations to this study, the first scenario is that the observed orbiting in the photosphere may be caused by one magnetic flux tube being physically moved around another by sub-photospheric flows with sufficient energy.
The second scenario is that the orbiting may be an apparent motion caused by the emergence of two intertwined sub-photospheric flux tubes.
\citet{min2009rotating} studied a case of sunspot rotation and they suggest that this was an apparent motion caused by the emergence of twisted flux tube, but were unable to rule out the effect of a torque force from the solar interior.
Contrarily, simulations by \citet{sturrock2015rotation} support the torque scenario for sunspot rotation and rule out the possibility of the rotation being an apparent effect.

To investigate whether the orbiting in this study is a product of flux emergence, the evolution of magnetic flux in each active region is examined.
The confined flares observed in the regions that produced the CMEs on 13 March 2012, 13 June 2012, and 14 June 2012 occurred when the active regions were in phases of flux emergence. 
In the case of the 8 October 2012 CME, the confined flares took place during a phase of decaying magnetic flux, and before the 14 July 2017 CME, confined flaring occurred during periods of both flux emergence and decay.
In the cases of all five CMEs, active region magnetic flux was increasing for at least 24 hours before eruption.

The lack of an observed increase in active region flux content during some of the confined flaring (coronal flux rope formation) does not necessarily rule out the scenario that the observed orbiting motions are a manifestation of an emerging flux tube.
If we consider a twisted $\Omega\textrm{-shaped}$ flux tube emerging through the photosphere, we would expect an increase in observed magnetic flux as the tube's apex rises into the solar atmosphere. However, once the apex has finished crossing the photospheric boundary, the mostly-vertical legs of the flux tube may continue to emerge with no discernible change in magnetic flux.
It is therefore difficult to comment on the origin of the observed orbiting motions in this study.
Regardless, the main conclusion is that magnetic reconnection occurred in the corona in every event examined here.

\section{Conclusions} \label{sec:sci3_conclusions}

Five hot flux ropes have been identified that formed via magnetic reconnection in the corona with estimated heights ranging from $\sim56\textrm{--}176\ \textrm{Mm}$, where the plasma $\beta < 1$.
The reconnection that built the flux ropes occurred in sporadic bursts, as evidenced by confined solar flares, and the timings of these flares suggests the flux ropes began to form from around 18--115 hours before they erupted.
Each confined flare event transformed sheared arcade field in to flux rope field and fed more flux to the growing structure.

In searching for the photospheric process(es) that caused magnetic reconnection to occur in the corona, it was found that all of the active regions exhibited newly-emerged magnetic flux fragments that moved towards and then orbited around pre-existing sunspots at some point during the time period of study.
The studied flux ropes each had one leg rooted in the sunspots where the strongest orbiting was observed, and there was consistency between the handedness of the emerging magnetic flux, the sense of orbiting, and the handedness of the flux ropes that formed.
Three left-handed sigmoids formed that were rooted in left-handed active regions that showed anticlockwise orbiting and two right-handed sigmoids formed in a right-handed active region that showed clockwise orbiting.
Furthermore, the event with the shortest amount of time between successive eruptions featured the strongest rate of orbiting, with $\approx 100\degr$ of motion measured between the CMEs on 13 and 14 June 2012 (although this is not observed to be a general correlation when considering the other regions studied).

It is inferred that the orbiting motions observed in the photosphere enable magnetic reconnection in the corona, forming hot magnetic flux ropes in HFT configurations above sheared, flaring arcades.
We propose that this mechanism be considered as a new trigger for CMEs that originate from hot flux ropes.

\begin{acknowledgements}
A.W.J. is supported by a European Space Agency (ESA) Research Fellowship, and acknowledges the support of the Leverhulme Trust and the Royal Society. 
L.M.G. is grateful to the Royal Society for the award of a Royal Society Research Fellowship that has supported this work.
L.v.D.G. is partially funded under STFC consolidated
grant number ST/N000722/1. L.v.D.G.  acknowledges
the Hungarian National Research, Development and Innovation Office grant
OTKA K-131508.
G.V. acknowledges the support of the Leverhulme Trust Research Project Grant 2014-051, the support from the European Union's
Horizon 2020 research and innovation programme under grant agreement No 824135, and of the STFC grant number ST/T000317/.
Data courtesy of NASA/SDO and the AIA and HMI science teams.
This research has made use of SunPy v1.1.0, an open-source and free community-developed solar data analysis Python package \citep{sunpy2020}.
\end{acknowledgements}

\bibliographystyle{aa}
\bibliography{bibliography}

\end{document}